\documentclass[preprint,review,12pt]{elsarticle}

\usepackage{amssymb}
\usepackage{amsthm}

\usepackage{bm}
\usepackage{amsmath}
\usepackage{epstopdf}
\usepackage{hyperref}
\usepackage{booktabs}
\usepackage{textcomp}
\usepackage{xcolor}
\usepackage{setspace}
\usepackage{geometry}
\usepackage{adjustbox}
\usepackage{rotating}

\graphicspath{{Fig/}}

\journal{}

\begin{document}

\begin{frontmatter}



\title{A predictive model of the turbulent burning velocity for planar and Bunsen flames over a wide range of conditions}



\author[fir,sec]{Zhen Lu}

\author[fir,sec,thi]{Yue Yang}
\ead{yyg@pku.edu.cn}

\address[fir]{State Key Laboratory for Turbulent and Complex Systems, College of Engineering, Peking University, Beijing 100871, China}
\address[sec]{BIC-ESAT, Peking University, Beijing 100871, China}
\address[thi]{CAPT, Peking University, Beijing 100871, China}

\begin{abstract}
We propose a predictive model of the turbulent burning velocity $s_T$ over a wide range of conditions.
The model consists of sub models of the stretch factor and the turbulent flame area.
%
The stretch factor characterizes the flame response of turbulence stretch and incorporates effects of detailed chemistry and transport with a lookup table of laminar counterflow flames.
The flame area model captures the area growth based on Lagrangian statistics of propagating surfaces, and considers effects of turbulence length scales and fuel characteristics.
%
The present model predicts $s_T$ via an algebraic expression without free parameters.
It is validated against 285 cases of the direct numerical simulation or experiment reported from various research groups on planar and Bunsen flames over a wide range of conditions, covering fuels from hydrogen to \textit{n}-dodecane, pressures from 1 to 20 atm, lean and rich mixtures, turbulence intensity ratios from 0.35 to 110, and turbulence length ratios from 0.5 to 80.
The comprehensive comparison shows that the proposed $s_T$ model has an overall good agreement over the wide range of conditions, with the averaged modeling error of 25.3\%.
Furthermore, the model prediction involves the uncertainty quantification for model parameters and chemical kinetics to extend the model applicability.
\end{abstract}



\end{frontmatter}


\section{Introduction}

The turbulent burning velocity (or the turbulent flame speed) $s_T$ is one of the most important statistics for turbulent premixed combustion~\cite{Peters_2000,Lipatnikov_2002,Driscoll_2008,Driscoll_2020}.
It is an indicator of the reactant consumption rate or a measure of the flame propagation speed, closely related to the fuel efficiency, heat release rate, and flame dynamics.
It is also used in various turbulent combustion models to close the nonlinear source term, as a crucial component in the modeling of turbulent premixed combustion.
Among problems related to $s_T$, a predictive model with a small set of key characteristic parameters is of practical interest for industrial design and combustion modeling. However, extensive studies have shown that $s_T$ depends on a variety of factors~\cite{Peters_2000,Lipatnikov_2002,Driscoll_2008,Driscoll_2020}, which poses an enormous challenge for developing a simple predictive model of $s_T$.


Various scaling laws and empirical models~\cite{Kobayashi_2005,Muppala_2005,Bradley_2013,Chaudhuri_2013,Wu_2015,Fogla_2017,Nguyen_2019} have been proposed for $s_T$ (also see review articles~\cite{Lipatnikov_2002,Driscoll_2008}) based on data of the direct numerical simulation (DNS) and experiment.
These algebraic models involve various flow parameters, dimensionless numbers, and model parameters for fitting the data.
The major issue is that the model parameters are sensitive to flame configurations, flow conditions, and the definition of $s_T$~\cite{Verma_2016}, requiring \emph{ad hoc} adjustment for different conditions.
The lack of a theoretical framework leads to the failure of the fitted correlations in a wide range of conditions, including turbulence intensity, integral length scale, reactant species, equivalence ratio, pressure, etc.

Theoretical development of the $s_T$ model is generally based on the flamelet concept~\cite{Peters_2000} and the flame area estimation~\cite{Damkohler_1940}.
For turbulent premixed flames at high Reynolds numbers, Zimont~\cite{Zimont_1979} derived a model by combining the small-scale turbulence effect on enhancing turbulent transport and the large-scale turbulence effect on wrinkling the flame surface.
The fractal theory was applied to estimate the flame area in turbulence and then to develop a variety of $s_T$ models~\cite{Gouldin_1987,Kerstein_1988d,Gulder_1991b}.
Another approach to model the flame area is via the $G$-equation~\cite{Williams_1985,Kerstein_1988c}.
Yakhot~\cite{Yakhot_1988} derived a model based on the dynamic renormalization group and the $G$-equation.
Peters~\cite{Peters_2000} obtained an algebraic expression through the balance of turbulent production, flame propagation, and scalar dissipation terms.
These models characterize the turbulence effects on $s_T$ through the flame area, but neglect the flame stretch, so they cannot predict fuel effects on $s_T$.

The flame stretch effect has been investigated numerically and experimentally~\cite{Hawkes_2006,Lapointe_2016,Savard_2017,Abbasi-Atibeh_2019,Lu_2020,Attili_2020b}, and this effect on flamelet speed is represented by the stretch factor $I_0$~\cite{Bray_1990}.
%
Diffusionally neutral flames have $I_0\approx 1$~\cite{Hawkes_2006}, while the thermal-diffusive effects cause $I_0$ away from unity~\cite{Lapointe_2016,Savard_2017,Abbasi-Atibeh_2019,Lu_2020} indicating a strong dependence of $s_T$ on flame stretch.
There are some efforts to incorporate the effects of detailed chemistry and transport in $s_T$ modeling~\cite{Bray_1990,Cant_1990,Bray_1991}.
A library of strained laminar flame was built to calculate $I_0$~\cite{Cant_1990}, and then an empirical model of $I_0$ with fitting experiment data was developed to reduce the use of pre-computed libraries~\cite{Bray_1990}.
%
Most existing models only treat either the flame area or the stretch effects on $s_T$. This limitation hinders a robust performance on the prediction of $s_T$ over a wide range of conditions.

Recently, You and Yang~\cite{You_2020} proposed a model of $s_T$ from the Lagrangian perspective.
This model predicts $s_T$ for flames of several simple fuels at 1 atm, with universal turbulence-related model constants based on Lagrangian statistics of propagating surfaces~\cite{Girimaji_1992,Zheng_2017} in non-reacting homogeneous isotropic turbulence~(HIT).
Later Lu and Yang~\cite{Lu_2020} investigated the lean hydrogen turbulent premixed flames at a range of pressures. They developed a model of the stretch factor $I_0$ with flamelet libraries to characterize the strong flame stretch effect on $s_T$ at high pressures.
Combining the models of flame area and stretch factor yields a predictive model of $s_T$, and this model was validated with the DNS result.
The model application is, however, restricted to a simple flame geometry, i.e., statistically planar flame propagation in HIT. Additionally, the ratio between the turbulence integral length and the flame thermal thickness is close to unity, and fuels are relatively simple such as hydrogen and methane in previous works~\cite{Lu_2020, You_2020}.

In the present study, we extend the Lagrangian-based modeling approach~\cite{Lu_2020, You_2020} to develop a predictive $s_T$ model for a wide range of conditions.
Several new modeling ingredients are added into the existing model of $s_T$, including a scaling of turbulence length scales for the effects of turbulence diffusivity and an empirical model of a fuel-dependent coefficient for instabilities and fuel chemistry.
%
%
This effort makes a crucial step towards a universal $s_T$ model for various turbulent flames without free parameters. We then use 285 DNS/experimental cases to assess the model performance.
The datasets from a number of research groups cover fuels from hydrogen to \textit{n}-dodecane, pressure from 1 to 20 atm, lean and rich mixtures, and a wide range of turbulence parameters. Flame configurations include planar and Bunsen flames, in accordance with the same consumption-based concept for model development.
In addition, it is inevitable to include empirical parameters in modeling of $s_T$ due to the many factors influencing $s_T$. The inclusion of flame stretch effects with detailed chemistry and transport also introduces uncertainty through the chemical kinetic model. We quantify the model uncertainty with respect to the parameters and chemical kinetic model to extend the model applicability.

The rest of this paper is organized as follows.
We present DNS/experimental cases of turbulent planar and Bunsen flames used for model development and assessment in Section~\ref{sec:cases}, and develop the predictive model of $s_T$ in Section~\ref{sec:model_dev}.
Section~\ref{sec:uq} discusses the uncertainty quantification on model predictions.
Comprehensive comparisons of model predictions against DNS/experimental data of $s_T$ are presented in Section~\ref{sec:assessment}.
Conclusions are drawn in Section~\ref{sec:conclusion}.

\section{DNS and experimental cases}
\label{sec:cases}

We collect a number of DNS and experimental datasets to develop and validate the model of $s_T$ for planar and Bunsen flames over a wide range of conditions.
Each dataset consists of a series of DNS/experimental cases for a fuel in one referenced paper.
The cases in a dataset are under various conditions, e.g., the pressure $p$, equivalence ratio $\phi$, unburnt temperature $T_u$, and turbulence intensity $u'$ and integral length scale $l_t$.
In particular, every case has a value of $s_T$ measured via either DNS or experiment.

\begin{figure}
  \centering
  \includegraphics[width=150 mm]{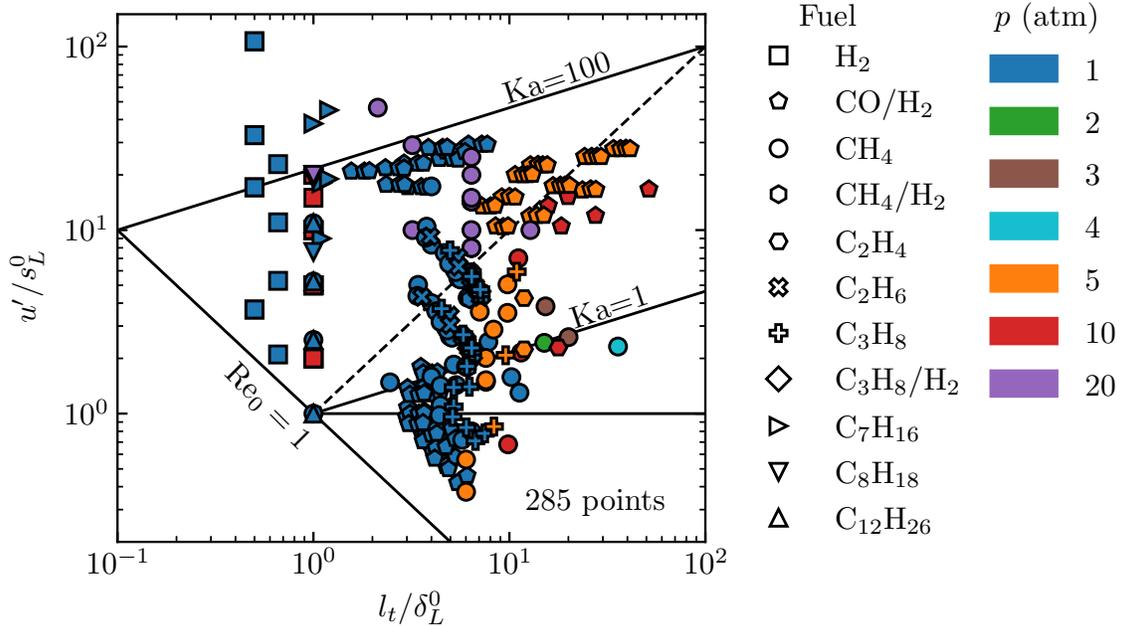}
  \caption{Parameters of DNS/experimental cases for model assessment in the regime diagram of turbulent premixed combustion. Each data point corresponds to one case, with different symbols for fuel species and colors for pressures.}
  \label{fig:diagram}
\end{figure}

Table~\ref{tab:cases} lists the datasets employed in the present study, together with the ranges of $p$, $\phi$, and $T_u$.
These datasets include 285 DNS/experimental cases from 17 papers from worldwide research groups, and cover a wide range of conditions. The fuel species varies from the hydrogen to large hydrocarbon molecules, with the pressure up to 20 atm and the equivalence ratio from very lean to rich.
Figure~\ref{fig:diagram} plots the parameters of the 285 cases in the diagram of turbulent premixed combustion.
Here, $\mathrm{Re}_0=\left(u'/s_L^0\right)\left(l_t/\delta_L^0\right)$ is the turbulent Reynolds number, $\mathrm{Ka}=\left(u'/s_L^0\right)^{\frac{3}{2}}\left(\delta_L^0/l_t\right)^{\frac{1}{2}}$ is the Karlovitz number, where $s_L^0$ and $\delta_L^0$ denote the laminar flame speed and thermal thickness of the unstrained one-dimensional laminar flame, respectively.
The scattered data points indicate a broad distribution of case parameters, with $u'/s_L^0$ from 0.35 to 110 and $l_t/\delta_L^0$ from 0.5 to 80.

\begin{sidewaystable}
    \centering
    \caption{DNS/experimental datasets used for model assessment of $s_T$.}
    \begin{tabular}{lccccc}
      \toprule[1.5pt]
      Datatset  &
      Configuration & Fuel              & $p$ (atm) & $\phi$      & $T_u$ (K)     \\
      \midrule[1.0pt]
      1. Aspden \textit{et al}., 2011~\cite{Aspden_2011}                          &
      planar        & H$_2$             & 1         & 0.31, 0.4   & 298           \\
      2. Aspden \textit{et al}., 2015~\cite{Aspden_2015}                          &
      planar        & H$_2$             & 1         & 0.4         & 298           \\
      3. Lu and Yang, 2020~\cite{Lu_2020}                                         &
      planar        & H$_2$             & 1-10      & 0.6         & 300           \\
      4. Aspden \textit{et al}., 2016~\cite{Aspden_2016}                          &
      planar        & CH$_4$            & 1         & 0.7         & 298           \\
      5. Aspden \textit{et al}., 2017~\cite{Aspden_2017}                          &
      planar        & CH$_4$            & 1         & 0.7         & 298           \\
      6. Wang \textit{et al}., 2017~\cite{Wang_2017}                              &
      planar        & CH$_4$            & 20        & 0.5         & 810           \\
      7. Lapointe \textit{et al}., 2015~\cite{Lapointe_2015}                      &
      planar        & C$_7$H$_{16}$     & 1         & 0.9         & 298, 500, 800 \\
      8. Savard \textit{et al}., 2017~\cite{Savard_2017}                          &
      planar        & C$_8$H$_{18}$     & 1, 20     & 0.9         & 298           \\
      9. Aspden \textit{et al}., 2017~\cite{Aspden_2017}                          &
      planar        & C$_{12}$H$_{26}$  & 1         & 0.7         & 298           \\
      10. Fragner \textit{et al}., 2015~\cite{Fragner_2015}                       &
      Bunsen        & CH$_4$            & 1-4       & 0.7-1.0     & 300           \\
      11. Muppala \textit{et al}., 2005~\cite{Muppala_2005}                       &
      Bunsen        & CH$_4$            & 1, 5, 10  & 0.9         & 298           \\
      12. Tamadonfar and G\"{u}lder, 2014~\cite{Tamadonfar_2014}                  &
      Bunsen        & CH$_4$            & 1         & 0.7-1.0     & 298           \\
      13. Tamadonfar and G\"{u}lder, 2015~\cite{Tamadonfar_2015}                  &
      Bunsen        & CH$_4$            & 1         & 0.7-1.35    & 298           \\
      14. Wable \textit{et al}., 2017~\cite{Wabel_2017}                           &
      Bunsen        & CH$_4$            & 1         & 0.75        & 298           \\
      15. Wang \textit{et al}., 2015~\cite{Wang_2015}                             &
      Bunsen        & CH$_4$            & 5, 10     & 1.0         & 298           \\
      16. Zhang \textit{et al}., 2018~\cite{Zhang_2018}                           &
      Bunsen        & CH$_4$            & 1         & 0.89        & 298           \\
      17. Venkateswaran \textit{et al},. 2015~\cite{Venkateswaran_2015}           &
      Bunsen        & CO/H$_2$          & 1, 5, 10  & 0.5-0.7     & 298           \\
      18. Zhang \textit{et al}., 2018~\cite{Zhang_2018}                           &
      Bunsen        & CO/H$_2$          & 1         & 0.5-0.7     & 298           \\
      19. Zhang \textit{et al}., 2020~\cite{Zhang_2020}                           &
      Bunsen        & CH$_4$/H$_2$      & 1         & 0.69-0.91   & 298           \\
      20. Muppala \textit{et al}., 2005~\cite{Muppala_2005}                       &
      Bunsen        & C$_2$H$_4$        & 5, 10     & 0.7         & 298           \\
      21. Tamadonfar and G\"{u}lder, 2015~\cite{Tamadonfar_2015}                  &
      Bunsen        & C$_2$H$_6$        & 1         & 0.7-1.45    & 298           \\
      22. Zhang \textit{et al}., 2018~\cite{Zhang_2018}                           &
      Bunsen        & C$_3$H$_8$        & 1         & 0.76        & 298           \\
      23. Tamadonfar and G\"{u}lder, 2015~\cite{Tamadonfar_2015}                  &
      Bunsen        & C$_3$H$_8$        & 1         & 0.8-1.35    & 298           \\
      24. Muppala \textit{et al}., 2005~\cite{Muppala_2005}                       &
      Bunsen        & C$_3$H$_8$        & 5         & 0.9         & 298           \\
      \bottomrule[1.5pt]
    \end{tabular}
    \label{tab:cases}
\end{sidewaystable}
%

In all the selected datasets, $s_T$ is computable based on the definition of the global consumption speed, which is an important criterion for our dataset selection.
%
It is noted that there are several definitions for $s_T$, i.e., the global consumption speed, local consumption speed, and local displacement speed~\cite{Driscoll_2008, Poinsot_2012}, and which definition to be employed depends on the flame configuration and measurement method.
%
%
Since the calculated value of $s_T$ may vary with its definition~\cite{Driscoll_2008}, the comparison of $s_T$ from the data and the model must be based on the same definition.

%

In general, there are two flame configurations for the datasets in Table~\ref{tab:cases}.
The configuration of DNS datasets is the statistically planar turbulent premixed flame propagating in HIT, which has been extensively studied for turbulence-flame interactions~\cite{Im_2016}.
The consumption speed in the DNS is calculated by the integration of the fuel consumption rate over the entire computational domain as~\cite{Poinsot_2012}
\begin{equation}
  s_T = \dfrac{1}{\rho_u A_L\left(Y_{\mathrm{F},b}-Y_{\mathrm{F},u}\right)} \int_{\Omega} \rho \dot{\omega}_{\mathrm{F}}dV,
\end{equation}
where $\rho_u$ is the density of unburnt mixture, $A_L$ is the flame surface area of the laminar flame, $Y_{\mathrm{F},u}$ and $Y_{\mathrm{F},b}$ are the mass fractions of fuel species in unburnt and burnt mixtures, respectively, $\Omega$ denotes the computational domain, and $\dot{\omega}_{\mathrm{F}}$ is the reaction rate of the fuel species. In Table~\ref{tab:cases}, all the DNS datasets are labeled by ``planar''.

For the experimental cases, the global consumption speed
\begin{equation}
  s_T = \dfrac{\dot{m}}{\rho_u A}
  \label{eq:st_exp}
\end{equation}
is calculated by the ratio of the total mass flow rate $\dot{m}$ of reactants and the averaged flame area $A$~\cite{Driscoll_2008}, where $A$ is calculated by an averaged progress variable $\langle c\rangle$ isocontour obtained with time-averaging of instantaneous flame images of radical signals.
%
It is necessary that all of the reactants pass through the flame brush in the experiment to calculate $s_T$ via Eq.~\eqref{eq:st_exp}.
For Bunsen flames, the burner exit is enveloped by the flame brush, satisfying the requirement.
A number of groups reported $s_T$ over a wide range of conditions in turbulent Bunsen flames, and these datasets are labeled by ``Bunsen'' in Table~\ref{tab:cases}.
%
Although various values of $\langle c\rangle$ from 0.05 to 0.5 were used~\cite{Kobayashi_2005,Tamadonfar_2015,Wabel_2017,Wang_2015},
$s_T$ data calculated with small $\langle c\rangle$ from 0.05 to 0.2 is employed in the present study in accordance with the turbulence parameters measured near the burner exit.

Moreover, there are several other flame geometries widely adopted for experimental measurements on $s_T$, such as V-shaped flames, counterflow flames, and spherical flames, but different definitions of $s_T$ were employed in these experiments.
In principle, the $s_T$ model should be validated against data obtained with the same definition~\cite{Driscoll_2008}, so the model assessment with different $s_T$ definitions is only briefly discussed in Section~\ref{sec:performance}.


%
%

\section{Model development}\label{sec:model_dev}

In the present modeling approach, we first loosely decouple the contributions of different processes to $s_T$, and model each process explicitly. Then, all the sub models are combined into a predictive model of $s_T$ without free parameters.

Utilizing the consumption-based definition of $s_T$ and the flamelet concept~\cite{Bray_1990}, the Damk\"{o}hler hypothesis~\cite{Damkohler_1940} suggests
\begin{equation}\label{eq:st_IA}
  \dfrac{s_T}{s_L^0} = I_0 \dfrac{A_T}{A_L}
\end{equation}
where $A_T$ denotes the turbulent flame area.
The form of Eq.~\eqref{eq:st_IA} implies that the turbulence influence on $s_T$ is decomposed into two parts, the stretch factor due to the flame response under flow variations~\cite{Bray_1990,Poinsot_2012}, and the flame area ratio due to the strain-rate and curvature effects in turbulence~\cite{Peters_2000,Driscoll_2008}.
In the present model, $I_0$ and $A_T/A_L$ in Eq.~\eqref{eq:st_IA} are modeled separately, and the influence of the flame stretch on the flame area is considered via the local flame speed in the modeling of $A_T/A_L$.
Next, we introduce the sub models involved in Eq.~\eqref{eq:st_IA} for predicting $s_T$.

\subsection{Stretch factor}

The flame stretch factor
\begin{equation}
  I_0 = \dfrac{\langle s_L\rangle_A}{s_L^0}
  \label{eq:I0_sL}
\end{equation}
characterizes the effect of chemical kinetics and molecular transport on $s_T$, and it links the mean laminar flamelet consumption speed $\langle s_L\rangle_A$ to the unstretched laminar flame speed $s_L^0$, where $\langle\cdot\rangle_A$ denotes the average over the flame surface.
The model of $I_0$ was proposed by Lu and Yang~\cite{Lu_2020} and is improved in the present study on the modeling of turbulence stretch.

As the first-order approximation, strain rate effects are neglected in many models by simply setting $I_0=1$ and $\langle s_L\rangle_A=s_L^0$ in Eq.~\eqref{eq:I0_sL}~\cite{Lipatnikov_2002,Driscoll_2008}.
%
On the other hand, $I_0$ plays an important role on $s_T$ in the cases with strong thermal-diffusive effects.
For instance, the lean hydrogen/air mixture with $\phi=0.6$ and $p=10\;\mathrm{atm}$ has $I_0\approx 3$ in strong turbulence~\cite{Lu_2020}.
Comparing with the case at $p=1$ atm, $s_T$ of lean hydrogen flames raises apparently with pressure because of the flame stretch effects.
Furthermore, the growth of $I_0$ with the turbulence intensity suppresses the bending of $s_T$ in strong turbulence.

In turbulent flames, $I_0$ depends on the distribution of the curvature and strain rate over the local flamelet.
The linear model of the laminar flame speed~\cite{Poinsot_2012} and symmetric distribution of flame curvature~\cite{Lu_2020,Lapointe_2015} suggest that effects of positive and negative flame curvatures tend to cancel out in strong turbulence, and the influence of the strain rate can be approximated with a presumed probability density function. Assuming the Dirac distribution for simplicity, the averaged consumption speed over the flame surface can be approximated as $\langle s_L\rangle_A=s_L$~\cite{Poinsot_2012}.

The thermal-diffusive effects alter $s_L$ with respect to the stretch on flames.
Asymptotic analysis~\cite{Poinsot_2012} showed a simple relation $s_L/s_L^0 = 1 - \mathrm{Ma Ka}$ for weak or moderate stretch,
where Ma is the Markstein number. 
In order to investigate the stretch effects with detailed chemistry and molecular transport, the response of $s_L$ to stretch in one-dimensional stretched flames, such as counterflow and cylindrical flames, can be employed as reference solutions~\cite{Cant_1990,Amato_2015}.

We model $I_0$ using a lookup table $\mathcal{F}$ formed by laminar flame data~\cite{Lu_2020} to capture the effects of detailed fuel chemistry and transport.
%
Laminar counterflow flames with two streams of the cold mixture and the corresponding equilibrium product are simulated to build the table $\mathcal{F}$.
For each counterflow flame solution, the consumption speed is calculated to obtain the ratio $s_L/s_L^0$, and the strain rate $a_t$ is estimated by the velocity gradient at the location with the maximum fuel consumption rate.
Using a series of counterflow simulations from weak stretched flames to extinction, a table of $s_L/s_L^0$ versus $\mathrm{KaLe}$ is obtained, where Le is the effective Lewis number of mixture, and the Karlovitz number of the laminar flame is calculated as
\begin{equation}
  \mathrm{Ka} = a_t \frac{\delta_L^0}{s_L^0}.
\end{equation}
For a certain condition, the stretch factor of the turbulent premixed flame is retrieved from the table as
\begin{equation}\label{eq:I0}
  I_0\left(K\right)
  = \dfrac{s_L\left(K\right)}{s_L^0}
  = \mathcal{F}\left(K\sqrt{\dfrac{p}{p_0}}\right).
\end{equation}
Here,
\begin{equation}\label{eq:K}
  K = 0.157 \left(\dfrac{u'}{s_L^0}\right)^2 \mathrm{Re}^{-\frac{1}{2}}
\end{equation}
is the model proposed by Bradley \emph{et al}.~\cite{Bradley_1992a,Bradley_1992b} for the turbulence stretch effect on flames, and $p_0=1\;\mathrm{atm}$ is a reference value for normalization, where $\textrm{Re} = u'l_t/\nu$ denotes the Reynolds number.
We find that this model is more generalized than that we used in Eq.~(7) in Ref.~\cite{Lu_2020}.

By incorporating the effects of detailed chemistry and transport, the $I_0$ model in Eq.~\eqref{eq:I0} is able to capture the response of $s_L$ to stretch for various reactants. In particular, a preliminary application of the modeled $I_0$ on lean hydrogen flames demonstrated that this type of model significantly improves the predication of $s_T$ from the simple model of $I_0=1$ at a broad range of pressure conditions~\cite{Lu_2020}.


\subsection{Turbulent flame area}


To estimate the flame surface ratio $A_T/A_L$ in Eq.~\eqref{eq:st_IA}, we apply the modeling approach developed by You and Yang~\cite{You_2020} based on theoretical analysis on Lagrangian statistics of propagating surfaces~\cite{Girimaji_1992,Zheng_2017}. The essence of this modeling framework is summarized below.

In Eq.~\eqref{eq:st_IA}, $A_T/A_L$ is approximated by the area ratio $A(t^*)/A_L$ of global propagating surfaces at a truncation time $t^*$, which signals that the characteristic curvature of initially planar propagating surfaces has reached the statistically stationary state in non-reacting HIT. This state resembles the statistical equilibrium state in combustion between the flame area growth due to turbulent straining and the area reduction due to flame self-propagation.

In the modeling of flame wrinkling in turbulence, the temporal growth of
\begin{equation}\label{eq:A}
  \dfrac{A(t^*)}{A_L}=\exp (\xi t^*)
\end{equation}
is approximated by an exponential function, where the constant growth rate
\begin{equation}\label{eq:xi}
  \xi = \mathcal{A} + \mathcal{B}s_{L0}^0 I_0^2
\end{equation}
is approximated by a linear model in terms of $I_0$ and the dimensionless laminar flame speed $s_{L0}^0 = s_L^0/s_{L,\textrm{ref}}$ normalized by a reference value $s_{L,\textrm{ref}} = 1$ m/s. The constants $\mathcal{A}=0.317$ and $\mathcal{B}=0.033$ are determined by fitting DNS data for Lagrangian statistics of propagating surfaces in non-reacting HIT.

Accounting for $t^*$ at limiting conditions of very weak ($u'/s_L^0\rightarrow 0$) and very strong ($u'/s_L^0\rightarrow \infty$) turbulence, theoretical analysis and data fit of Lagrangian statistics of propagating surfaces yield
\begin{equation}\label{eq:t}
  t^* = T_\infty^*
    \left[1 - \exp
      \left(
        - \dfrac{\mathcal{C}\mathrm{Re}^{-1/4}}{\xi T_\infty^*}
        \dfrac{u'}{s_L^0 I_0}
      \right)
    \right],
\end{equation}
where $T_\infty^*=5.5$ is a universal truncation time describing the stationary state of material surfaces.
%
Substituting Eqs.~\eqref{eq:xi} and \eqref{eq:t} into Eq.~\eqref{eq:A} yields
\begin{equation}\label{eq:AT}
  \dfrac{A_T}{A_L} = \exp\left\{T_\infty^*\left(\mathcal{A}+\mathcal{B}s_{L0}^0 I_0^2\right)\left[1\!-\!\exp\left(-\dfrac{\mathcal{C}\,\mathrm{Re}^{-1/4}}{T_\infty^*\left(\mathcal{A}+\mathcal{B}s_{L0}^0 I_0^2\right)}\dfrac{u^\prime}{s_L^0 I_0}\right)\right]\right\},
\end{equation}
where turbulence related constants $\mathcal{A}$, $\mathcal{B}$, and $T_\infty^*$ are universal,
%
and $\mathcal{C}$ is a fuel-dependent coefficient which will be discussed later.

The flame area model in Eq.~\eqref{eq:AT} captures the dependence of $A_T$ on $u'$ in very weak and strong turbulence~\cite{Lu_2020,You_2020}.
As $u'\rightarrow 0$, the Taylor expansion of Eq.~\eqref{eq:AT} predicts a linear growth of $A_T$ with $u'$. As $u'\rightarrow \infty$, the modeled $A_T$ reaches an asymptotic state, showing the bending phenomenon of $s_T$ with $I_0\approx1$.
At $p=1$ atm, the validation with several DNS datasets showed that the model in Eq.~\eqref{eq:AT} captures the variation of $s_T$ with $u'$, outperforming previous models~\cite{You_2020}.

\subsection{Turbulence length scales}
\label{sec:length}

It is well recognized that $s_T$ has a dependence on length scales for turbulent premixed flames in the thin reaction zone~\cite{Attili_2020b,Wabel_2017,Nivarti_2019}, but this dependence is only implicitly characterized by the Reynolds number in Eq.~\eqref{eq:AT}.
In this work, we improve the area ratio model Eq.~\eqref{eq:AT} by introducing a scaling of the length scales and keeping the merit of Eq.~\eqref{eq:AT} at the two turbulence limits.

In small-scale turbulence, Damk\"{o}hler argued that turbulence affects scalar mixing only via turbulent transport.
In analogy to the scaling relation $s_L^0\sim \sqrt{D}$ of the laminar burning velocity and the molecular diffusivity $D$~\cite{Poinsot_2012}, the turbulent burning velocity is assumed to be proportional to the square root of turbulent diffusivity $D_T \sim u'l_t$ as
\begin{equation}\label{eq:l_scaling}
  \dfrac{s_T}{s_L^0} \sim \sqrt{\dfrac{D_T}{D}} \sim \sqrt{\dfrac{u'l_t}{s_L^0\delta_L^0}}.
\end{equation}

To have a dependence on $l_t/\delta_L^0$ as $u'/s_L^0\rightarrow 0$ in very weak turbulence, Eq.~\eqref{eq:t} is modified as
\begin{equation}\label{eq:l1}
  t^* = T_\infty^*
    \left\lbrace
      1 - \exp
      \left[
        - \dfrac{\mathcal{C}\mathrm{Re}^{-1/4}}{\xi T_\infty^*}
        \dfrac{u'}{s_L^0 I_0}
        \left(\dfrac{l_t}{\delta_L^0}\right)^{1/2}
      \right]
    \right\rbrace,
\end{equation}
and then the Taylor expansion of Eq.~\eqref{eq:A} for $u'\rightarrow 0$ becomes
\begin{equation}\label{eq:expansion}
  \dfrac{A_T}{A_L} = 1 + \mathcal{C}\mathrm{Re}^{-1/4}
                         \left(\dfrac{l_t}{\delta_L^0}\right)^{1/2}
                         \dfrac{u'}{s_L^0 I_0}
                   = 1 + \mathcal{C}
                         \mathrm{Re}^{-1/4}_F
                         \mathrm{Da}^{1/4}
                         \dfrac{u'}{s_L^0 I_0},
\end{equation}
where $\mathrm{Re}_F = s_L^0\delta_L^0/\nu$ is the flame Reynolds number with the kinematic viscosity $\nu$, $\mathrm{Da}=\left(l_t/\delta_L^0\right)\left(s_L^0/u'\right)$ is the Damk\"{o}hler number.
Equation~\eqref{eq:expansion} has the same scaling $s_T\sim \mathrm{Da}^{1/4}u'$ in the Zimont model~\cite{Zimont_1979}.
As $u'/s_L^0\rightarrow \infty$ in very strong turbulence, Eq.~\eqref{eq:AT} becomes $A_T/A_L=\exp(\xi T_\infty^*)$ and it is a constant for constant $I_0$. To match the scaling in Eq.~\eqref{eq:l_scaling}, we propose
\begin{equation}\label{eq:l2}
  \dfrac{A_T}{A_L} = \exp\left(\xi T_\infty^*\right)
                     \left(\dfrac{l_t}{\delta_L^0}\right)^{1/2}
                   = \exp
                   \left[
                     \xi T_\infty^*
                     + \dfrac{1}{2}\ln\left(\dfrac{l_t}{\delta_L^0}\right)
                   \right].
\end{equation}

Combining the corrections in Eqs.~\eqref{eq:l1} and \eqref{eq:l2}, Eq.~\eqref{eq:AT} is improved by considering the scaling with length scales as
\begin{equation}\label{eq:A*}
    \dfrac{A_T}{A_L} = \exp\left\lbrace\!\!\left[T_\infty^*\!\left(\mathcal{A}\!+\!\mathcal{B}s_{L0}^0 I_0^2\right)\!+\!{\dfrac{1}{2}\ln\left(\dfrac{l_t}{\delta_L^0}\right)}\right]\!\!\left[\!1\!-\!\exp\left(-\dfrac{\mathcal{C} \,\mathrm{Re}^{-1/4}\left(l_t/\delta_L^0\right)^{1/2}}{T_\infty^*\left(\mathcal{A}\!+\!\mathcal{B}s_{L0}^0 I_0^2\right)}\dfrac{u^\prime}{s_L^0 I_0}\right)\!\right]\!\!\right\rbrace.
\end{equation}
We remark that the Taylor expansion of Eq.~\eqref{eq:A*} in very weak turbulence is not exactly the same as Eq.~\eqref{eq:expansion} due to the factor of $\ln(l_t/\delta_L^0)/2$ introduced for strong turbulence in Eq.~\eqref{eq:l2}, whereas the scaling $s_T\sim\mathrm{Da}^{1/4}u'$ in Eq.~\eqref{eq:expansion} is kept.
In addition, Eq.~\eqref{eq:A*} leads to a questionable prediction $A_T/A_L < 1$ at the limit $l_t/\delta_L^0\rightarrow 0$, but this modeling defect only happens as $l_t/\delta_L^0 < \exp(-2\xi T_\infty^*) = 0.03$.
Regarding to the length scale $l_t/\delta_L^0 \geq O(1)$ in practical cases in Fig.~\ref{fig:diagram}, this shortcoming can be neglected for most applications.

We illustrate the importance of the length scale modeling in $A_T/A_L$ using three sets of DNS of planar flames~\cite{Lee_2010}.
It is noted that these DNS cases based on the progress variable are not included in Table~\ref{tab:cases} and Fig.~\ref{fig:diagram}.
We approximate $I_0=1$ for Le $=1$ for this DNS, and then Eq.~\eqref{eq:st_IA} is simplified to $s_T/s_L^0 = A_T/A_L$.
The cases are set to have the gas expansion $\rho_u/\rho_b=6$, where $\rho_b$ is the density of the burnt gas.
By adjusting the one-step chemistry coefficient, $s_L^0$ and $\delta_L^0$ in these cases are varied, as listed in Table~\ref{tab:Lee}.
%

\begin{table}
    \centering
    \caption{DNS parameters in Ref.~\cite{Lee_2010}.}
    \begin{tabular}{ccccc}
      \toprule[1.5pt]
      Group & $l_t$ (cm)  & $\delta_L^0$ (cm) & $s_L^0$ (cm/s)  & $l_t/\delta_L^0$  \\
      \midrule[1.0pt]
      R     & 0.65        & 0.22              & 0.300           & 2.955             \\
      T     & 0.65        & 0.40              & 0.163           & 1.625             \\
      L     & 0.34        & 0.22              & 0.300           & 1.545             \\
      \bottomrule[1.5pt]
    \end{tabular}
    \label{tab:Lee}
\end{table}

\begin{figure}
  \centering
  \includegraphics[width=80 mm]{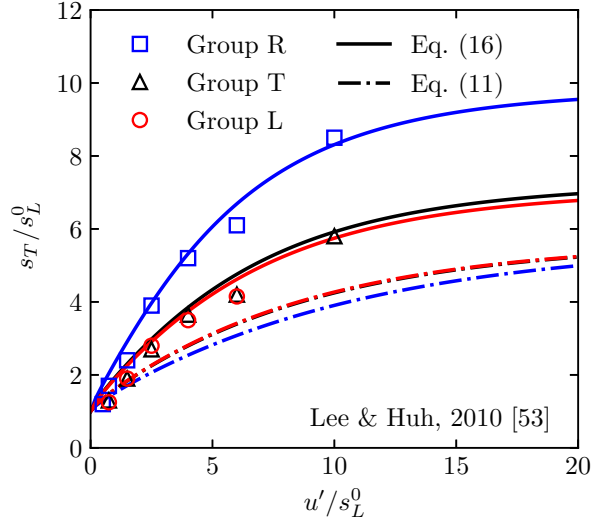}
  \caption{Comparison of $s_T$ calculated from the DNS~\cite{Lee_2010} (symbols), the model Eq.~\eqref{eq:A*} with length scale effects (solid lines), and the model Eq.~\eqref{eq:AT} without length scale effects (dash-dotted lines).}
  \label{fig:lt}
\end{figure}

Figure~\ref{fig:lt} compares the predictions of $s_T$ from the present model Eq.~\eqref{eq:A*} with the length scale effects and from the previous model Eq.~\eqref{eq:AT}, where $\mathcal{C}=0.83$ is further modeled by Eq.~\eqref{eq:C0f} with Le $=1$ and $I_0=1$. 
%
In groups R and L of this DNS series, the laminar flame parameters are the same, while $l_t/\delta_L^0= 2.955$ and 1.545 are different.
We find that the model prediction from Eq.~\eqref{eq:A*} (solid lines) agrees well with the DNS results (symbols), showing the growth of $s_T/s_L^0$ with $l_t/\delta_L^0$.
By contrast, the model in Eq.~\eqref{eq:AT} (dash-dotted lines) fails to predict different $s_T/s_L^0$ with the length scale effect in these groups.
In groups T and L, the laminar flame parameters are different and length scale ratios are close. Additionally, since $s_L^0$ is small in these groups, the contribution from $\xi$ in Eq.~\eqref{eq:xi} is negligible. Thus, the model predictions from Eq.~\eqref{eq:A*} are close for these two groups.

\subsection{Fuel-dependent coefficient}

In Eq.~\eqref{eq:A*}, the model coefficient $\mathcal{C}$ characterizes the fuel effect on the growth of $A_T$ and $s_T$ in weak turbulence with $u'/s_L^0 = O(1)$.
Since the hydrodynamic and thermal-diffusive instabilities drive the growth of $A_T$ in weak turbulence~\cite{Creta_2020}, unstable flames with Le $<$ 1 have large $\mathcal{C}$, while stable flames with Le $>1$ have small $\mathcal{C}$.
%
%
%
%
One way to determine $\mathcal{C}$ is from one or a few available DNS or experimental data points of $s_T$ in weak turbulence.
Alternatively, the constant value $\mathcal{C} = 2.0$ is suggested for hydrogen mixtures, and $\mathcal{C} =1.0$ is recommended for other fuels such as methane~\cite{You_2020}.

\begin{figure}
  \centering
  \includegraphics[width=80 mm]{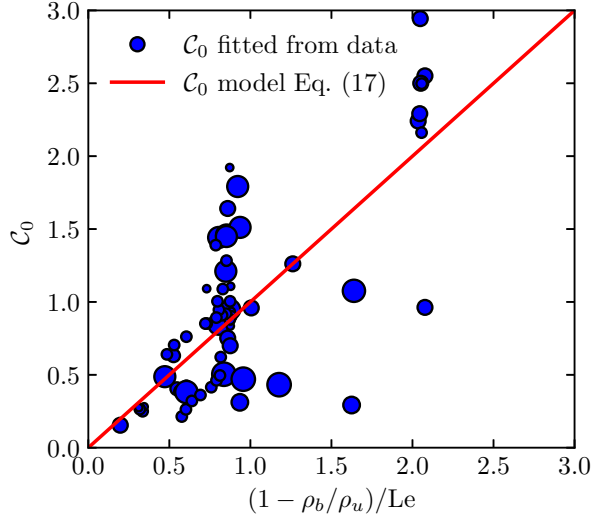}
  \caption{
    Fit of $\mathcal{C}_0$ in Eq.~\eqref{eq:C0f} using the DNS/experimental cases listed in Table~\ref{tab:cases}. Each marker represents one set of cases, and the marker size is proportional to the number of cases in the dataset.
    }
  \label{fig:C0}
\end{figure}

Towards a universal predictive model of $s_T$, we reduce the degree of freedom on the determination of $\mathcal{C}$ in the present work.
First, we decompose $\mathcal{C} = \mathcal{C}_0 I'_0$, where $I'_0=I_0(K=1)$ is obtained from stretch factor table $\mathcal F$, and $\mathcal{C}_0$ only depends on the mixture composition.
Figure~\ref{fig:C0} shows the fit of $\mathcal{C}_0$ from the DNS and experimental cases listed in Table~\ref{tab:cases} against the Lewis number of mixtures, where $\mathcal{C}_0$ is calculated using the nonlinear least square fit~\cite{SciPy_2020} for each case. The size of each symbol in Fig.~\ref{fig:C0} is proportional to the number of data points in the corresponding dataset. Note there is an apparent trend that $\mathcal{C}_0$ decreases with the increase of Le, though the data points are very scattered.

%
From the data fit of $\mathcal{C}_0$ in Fig.~\ref{fig:C0}, we propose an empirical model
\begin{equation}\label{eq:C0f}
  \mathcal{C}_0=\dfrac{1-\rho_b/\rho_u}{\mathrm{Le}}
\end{equation}
for various mixtures.
This model implies stronger hydrodynamic and thermal-diffusive instabilities at smaller $\rho_b/\rho_u$ and Le, respectively, and predicts large $\mathcal{C}_0$ for unstable cases.
%
We remark that the severe scattering points of $\mathcal{C}_0$ in Fig.~\ref{fig:C0} suggest that $\mathcal{C}_0$ may depend on multiple parameters rather than a simple function in Eq.~\eqref{eq:C0f}, so the data-driven methods can be used to fit $\mathcal{C}_0$ in the future work.
%
%
%

\subsection{Predictive model of $s_T$}

Substituting Eqs.~\eqref{eq:I0} and \eqref{eq:A*} with all the model constants into Eq.~\eqref{eq:st_IA}, we have a predictive model of the turbulent burning velocity
\begin{equation}\label{eq:st}
    \dfrac{s_T}{s_L^0} \!=\!
    \mathcal{F}
    \exp\!
      \left\lbrace\!\!
        \left[\!
          \left(
            1.742\!+\!0.182 s_{L0}^0 \mathcal{F}^2
          \right)
          \!+\!
          {\dfrac{1}{2}\ln\left(\!\dfrac{l_t}{\delta_L^0}\!\right)}
        \!\right]
        \!\!\!
        \left[\!
          1
          \!-\!
          \exp\left(\!\!
            -\dfrac{\left[1-\dfrac{\rho_b}{\rho_u}\right]\mathrm{Le}^{-1} \mathcal{F}'\,\mathrm{Re}^{-\frac{1}{4}}\left[\dfrac{l_t}{\delta_L^0}\right]^\frac{1}{2}}{\left(1.742\!+\!0.182 s_{L0}^0 \mathcal{F}^2\right)\mathcal{F}}
            \dfrac{u^\prime}{s_L^0}
          \!\!\right)
        \!\right]
      \!\!\right\rbrace.
\end{equation}
In Eq.~\eqref{eq:st}, the model constants related to turbulence are universal, and the fuel-dependent parameters are obtained by the lookup table from a separate laminar flame calculation or the fit of DNS and experimental data in the literature.
Therefore, the model Eq.~\eqref{eq:st} has no free parameters.

From the algebraic model in Eq.~\eqref{eq:st}, $s_T$ is obtained from a given set of reactant and flow parameters. Specifically, the required inputs for $s_T$ predictions are the reactant species, equivalence ratio $\phi$, unburnt temperature $T_u$, pressure $p$, turbulence intensity $u'$, and integral length $l_t$.
Moreover, the free propagation of laminar premixed flames and a series of counterflow flames to extinction are calculated to obtain the laminar flame parameters, including laminar flame speed $s_L^0$, flame thermal thickness $\delta_L^0$, stretch factor table $\mathcal{F}$, and the Lewis number Le. Finally, $s_T/s_L^0$ is calculated by substituting $u'$ and $l_t$ into Eq.~\eqref{eq:st}.
The above procedure has been implemented by a modularized code available at \url{https://github.com/YYgroup/STmodel}.

\section{Uncertainty quantification}
\label{sec:uq}

The model in Eq.~\eqref{eq:st} gives an explicit and deterministic prediction of $s_T$ without free parameters, but this model and even the DNS/experimental results of $s_T$ are inevitably suffered by various uncertainties.
The possible uncertainties in the modeling come from model parameters, chemical kinetics, input data, etc.
%
Therefore, the uncertainty of model prediction should be quantified from the various sources~\cite{Smith_2013,Maitre_2010}, which is an essential supplement to a single deterministic model calculation from Eq.~\eqref{eq:st}. %


\subsection{Uncertainty from model parameters}
\label{sec:uq_para}

As mentioned in Section~\ref{sec:model_dev}, model parameters $\mathcal{A}$, $\mathcal{B}$, and $T_\infty^*$ are fitted from DNS data of nonreacting HIT \cite{You_2020}, the fuel-dependent coefficient $\mathcal{C}_0$ is calculated by an empirical expression fitted from a number of combustion DNS/experimental cases. The uncertainty in the fits causes the uncertainty of the model parameters.

\begin{figure}
  \centering
  \includegraphics[width=80 mm]{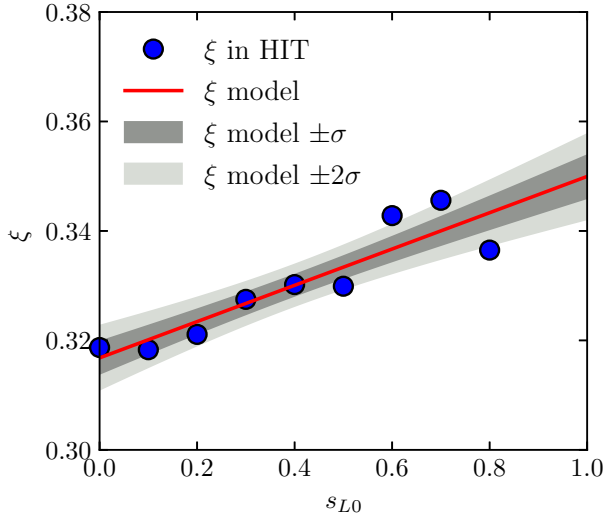}
  \caption{
    Uncertainty quantification in the modeling of $\xi$ in Eq.~\eqref{eq:xi}.
    Dark and light shades denote one and two standard deviations for the uncertainty range, respectively.
    }
  \label{fig:uncertainty}
\end{figure}

As an illustrative example, Fig.~\ref{fig:uncertainty} shows the uncertainty in the modeling of $\xi$ in Eq.~\eqref{eq:xi}. The uncertainty is obtained via the Bayesian linear regression~\cite{Smith_2013,pymc3} from Lagrangian statistics of propagating surfaces in nonreacting HIT with constant $s_{L0}$ and $I_0 =1$.
In this uncertainty quantification of model predictions, samples on $\mathcal{A}$ and $\mathcal{B}$ are generated with the Markov chain Monte--Carlo method~\cite{Smith_2013,pymc3}.
%
In Fig.~\ref{fig:uncertainty}, the solid line represents Eq.~\eqref{eq:xi}, and the dark and light shades denote the 68\% ($\pm\sigma$) and 95\% ($\pm 2\sigma$) confidence intervals on the $\xi$ model, where $\sigma$ denotes the standard deviation of model predictions of $\xi$ .
We observe that the fit of turbulence statistics can lead to a range of possible parameter values.

Similar to $\mathcal{A}$ and $\mathcal{B}$, other model parameters in Eq.~\eqref{eq:st} are associated with uncertainties.
In particular, the fit of $\mathcal{C}_0$ involves a large uncertainty regarding the severe scatter of data points shown in Fig.~\ref{fig:C0}.
In subsequential validations, the model parameters $\mathcal{C}_0$ and $T_\infty$ are presumed to satisfy the independent normal distribution $\mathcal{N}(\mu, \sigma)$ with the mean $\mu$ and the standard deviation $\sigma$,  and the Monte--Carlo method is applied to generate samples for these two parameters.
Table~\ref{tab:uncertainty} lists the uncertainty level $\sigma/\mu$ for the model parameters in Eq.~\eqref{eq:st}, which is specified based on sample statistics from DNS datasets.
%
%
The model prediction of $s_T$ is calculated for each sample, and then the uncertainty is calculated from the statistics of $s_T$.

\begin{table}
    \centering
    \caption{Setting of the uncertainty level for model parameters.}
    \begin{tabular}{ccc}
      \toprule[1.5pt]
      Parameter       & $\mu$                   & $\sigma/\mu$ \\
      \midrule[1.0pt]
      $\mathcal{A}$   & 0.317                   & 0.01 \\
      $\mathcal{B}$   & 0.033                   & 0.17 \\
      $T_{\infty}^*$  & 5.5                     & 0.10 \\
      $\mathcal{C}_0$ & ($1-\rho_b/\rho_u$)/Le  & 0.15 \\
      \bottomrule[1.5pt]
    \end{tabular}
    \label{tab:uncertainty}
\end{table}

\subsection{Uncertainty from chemical kinetics}
\label{sec:uq_chem}

The uncertainty in chemical kinetic models is typically associated to the reaction rate coefficient $k$~\cite{WangH_2015}. It propagates to the model prediction of $s_T$ via $s_L^0$, $\delta_L^0$, and $I_0$ that characterizes effects of detailed chemical kinetics and molecular transport in Eq.~\eqref{eq:st}.
In our implementation, one-dimensional freely propagating flames and counterflow flames are calculated to obtain $s_L^0$, $\delta_L^0$, and $I_0$. Consequently, the uncertainty of chemical kinetic models firstly propagates to the result of laminar flames.

In the uncertainty quantification of chemical kinetic models~\cite{WangH_2015}, the reaction rate coefficient $k_i$ of the $i$-th reaction is normalized into a factorial variable~\cite{Sheen_2009,Sheen_2011}
\begin{equation}
  x_i = \dfrac{\ln k_i/k_{i,0}}{\ln f_i},
\end{equation}
where $k_{i,0}$ is the nominal value of $k_i$, and $f_i$ is the uncertainty factor of the $i$-th reaction.
It is assumed that $x_i$ of all reaction coefficients satisfies the independent normal distribution $\mathcal{N}(\mu = 0, \sigma =1)$.

\begin{figure}
  \centering
  \includegraphics[width=80 mm]{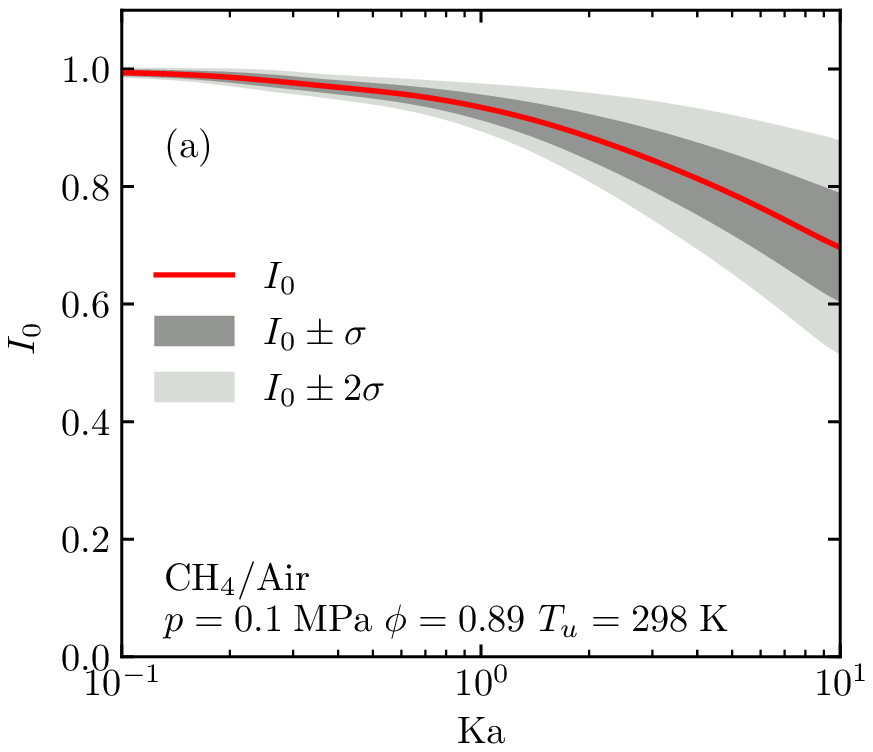}
  \includegraphics[width=80 mm]{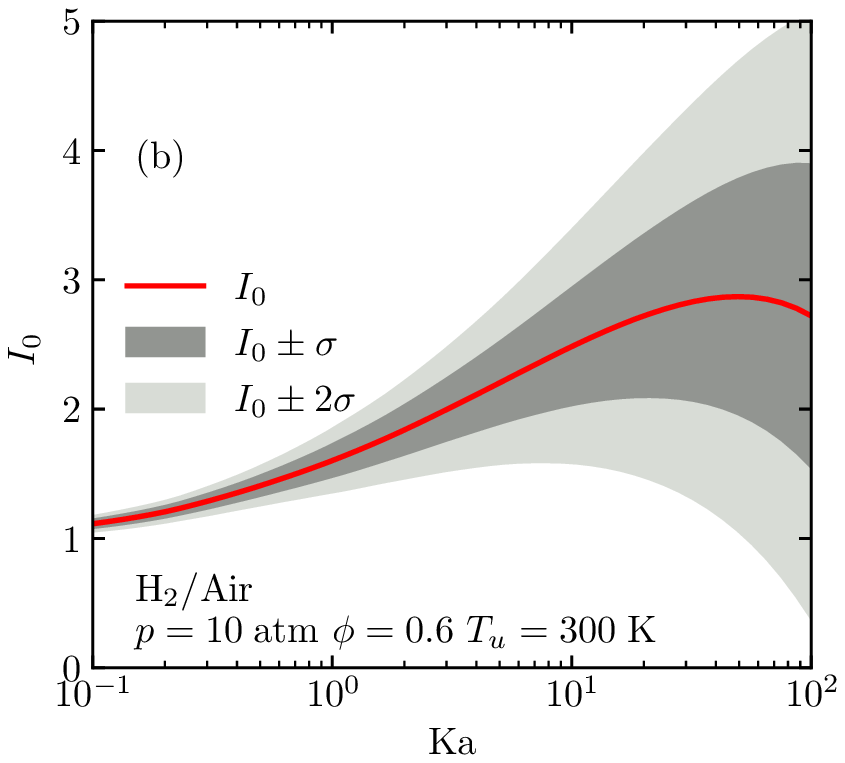}
  \caption{
    Uncertainty ranges for $I_0$ of (a) methane flames~\cite{Zhang_2018} and (b) hydrogen flames~\cite{Lu_2020}. Dark and light shades denote one and two standard deviations ($\sigma$) for the uncertainty range, respectively.
    }
  \label{fig:chemuq}
\end{figure}

We conducted chemical kinetic uncertainty quantification for the cases of hydrogen and methane flames with the FFCM-1 mechanism~\cite{Smith_FFCM}.
The uncertainty factors $f_i$ for kinetic rates are taken from the FFCM-1 model~\cite{Smith_FFCM,Tao_2018}.
The laminar flame parameters $s_L^0$, $\delta_L^0$, and $I_0$ for each sample are calculated with the Monte--Carlo sampling~\cite{Robert_1999} of $x_i$,
and are then used in Eq.~\eqref{eq:st} to propagate the chemical kinetics uncertainty to the $s_T$ prediction.
Uncertainties of $I_0$ for two cases in Refs.~\cite{Lu_2020,Zhang_2018} are shown in Fig.~\ref{fig:chemuq} for example.
We observe that the uncertainty range depends on many factors such as the fuel, equivalence ratio, and pressure.
In general, the flame speed exhibits large sensitivity on reaction rate coefficients at high pressures and near extinction conditions, so the uncertainty range increases with pressure and Ka.
For the hydrogen cases, the large chemical uncertainties of $I_0$, $s_L^0$, and $\delta_L^0$ lead to 95\% confidence intervals larger than 10 times of the predicted $s_T$, so it is not presented in the further model assessment for clarity.
By contrast, the uncertainty range for the methane flames is relatively small, and it is similar for other methane cases.

%
%

\section{Model assessment}
\label{sec:assessment}

We validate the model of $s_T$ in Eq.~\eqref{eq:st} using the DNS and experimental datasets listed in Table~\ref{tab:cases}.
The model predictions and the DNS/experimental results are first compared for each dataset, and the comparisons are categorized into different fuel types. Then, an overall model performance is assessed considering all the datasets.

In the implementation of the model, a separate simulation of the laminar counterflow flame is carried out for each case with the corresponding case conditions to obtain the stretch factor table $\mathcal{F}$. One-dimensional free flame simulations are conduced to obtain $s_L^0$ and $\delta_L^0$.
For each DNS case, the same chemical mechanism in the original DNS is applied for the calculation. 
For experimental cases, the FFCM-1 mechanism~\cite{Smith_FFCM} is applied for methane related cases, the Davis mechanism~\cite{Davis_2005} is used for syngas cases, the UCSD mechanism~\cite{UCSD} is employed for ethane, ethylene, and propane cases.
%
%
When the uncertainty range is presented in following figures, grey shades denote the uncertainty ranges due to the model parameter uncertainties discussed in Section~\ref{sec:uq_para}. Dark and light grey shades represent one and two standard deviations, corresponding to 68 and 95 percentage of confidence intervals, respectively.

\subsection{Hydrogen}

Hydrogen is a promising fuel for clean combustion applications without carbon emission.
The lean hydrogen flame at a negative Markstein number is thermal-diffusive unstable, and it has $I_0>1$ growing with pressure, so its turbulent burning velocity depends on both $I_0$ and $A_T/A_L$ in Eq.~\eqref{eq:st_IA}.
In particular, the growth of $I_0$ due to the turbulence stretch on flames can significantly enhance $s_T$ at high pressures~\cite{Lu_2020}.

\begin{figure}
  \centering
  \includegraphics[width=80 mm]{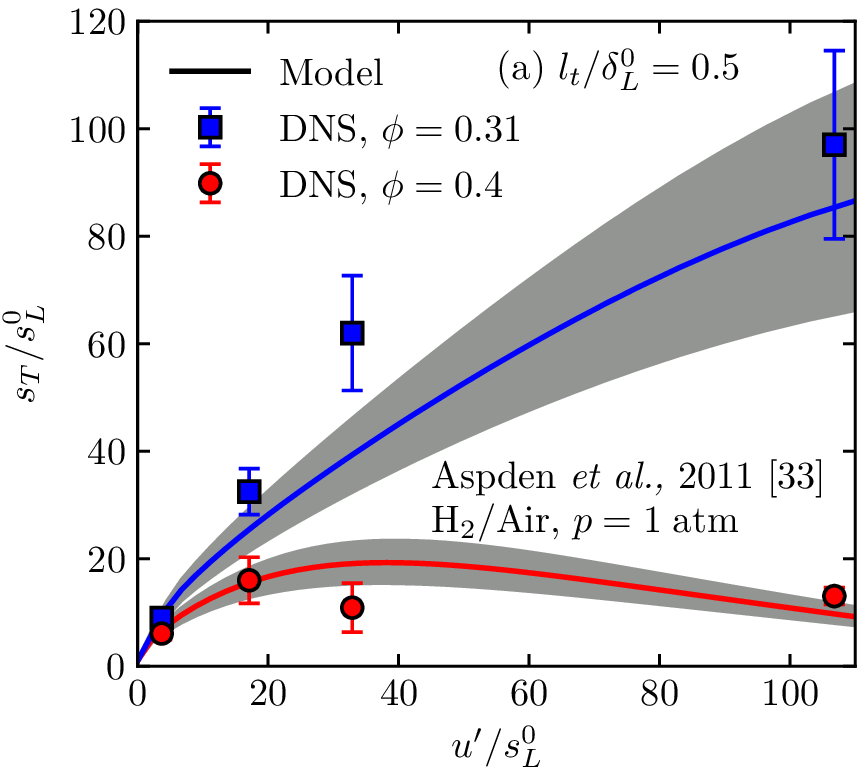}
  \includegraphics[width=80 mm]{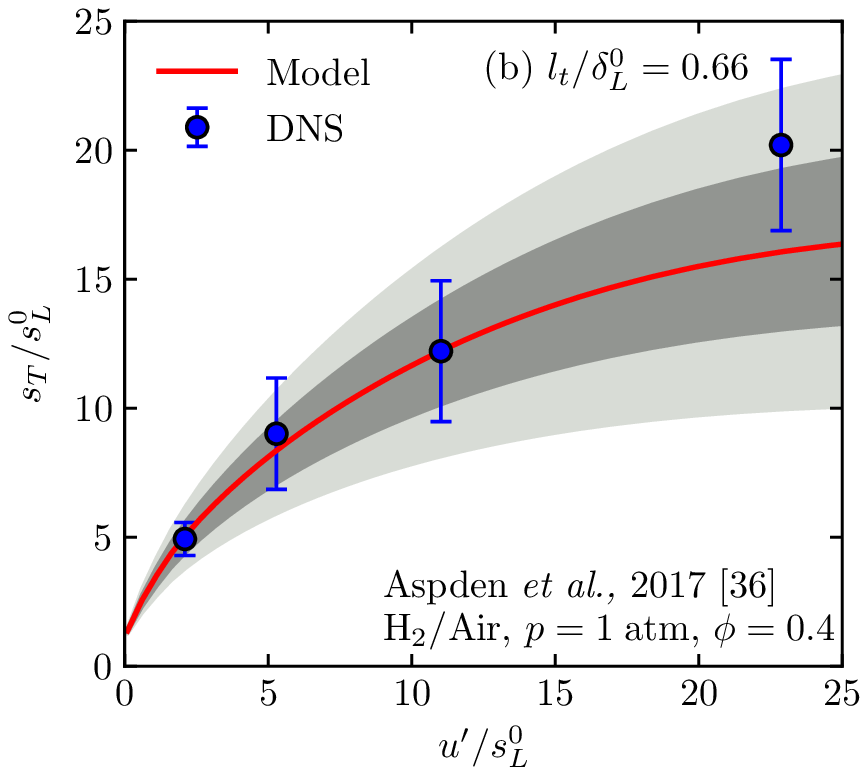}
  \caption{
    Comparisons of $s_T$ obtained from the DNS (symbols with error bars for one standard deviation) and the proposed model Eq.~\eqref{eq:st} (lines) for lean hydrogen flames at $p=1$ atm.
    (a) Cases~\cite{Aspden_2011} with $\phi=0.31$ and 0.4, and $l_t/\delta_L^0=0.5$; (b) Cases~\cite{Aspden_2017} with $\phi=0.4$ and $l_t/\delta_L^0=0.66$.
    Dark and light shades denote one and two standard deviations for the model uncertainty range, respectively.
    Only the shades for one standard deviation are presented in (a) for clarity.
    }
  \label{fig:h2_phi}
\end{figure}

Aspden \textit{et al}.~\cite{Aspden_2011,Aspden_2015,Aspden_2016,Aspden_2017} conducted a series of DNS on statistically planar turbulent premixed flames with various fuels at a wide range of turbulence intensities and corresponding Karlovitz numbers.
For the lean hydrogen flames, $s_T$ at different equivalence ratios and length scales was reported.
Figure~\ref{fig:h2_phi} compares the model predictions (lines) and DNS results (symbols) of $s_T$.
The DNS cases in Fig.~\ref{fig:h2_phi}a have two equivalence ratios $\phi=0.31$ and $\phi=0.4$ and the same length scale ratio $l_t/\delta_L^0=0.5$~\cite{Aspden_2011} with the GRIMech 2.11~\cite{Bowman_GRI}, and the cases in Fig.~\ref{fig:h2_phi}b have $\phi=0.4$ with a larger $l_t/\delta_L^0=0.66$~\cite{Aspden_2017} with the Li mechanism~\cite{Li_2004}.

As shown in Fig.~\ref{fig:h2_phi}a, the turbulent burning velocity grows almost linearly with the turbulence intensity for the very lean case with $\phi=0.31$ (blue squares), even at a very large $u'/s_L^0$ up to 100.
By contrast, the variation of $s_T$ with $u'$ shows a typical bending phenomenon for $\phi=0.4$ (red circles), where $s_T$ stops growing and then decays with $u'$ in strong turbulence.
The present model (solid lines) well captures the different trends for the two equivalence ratios, which is contributed by the variation of $I_0$ in Eq.~\eqref{eq:st} with laminar flame parameters. 
Figure~\ref{fig:h2_phi}b validates the model for a larger $l_t/\delta_L^0=0.66$~\cite{Aspden_2017}.
With the length scale effects incorporated in the modeling of $A_T$ in Eq.~\eqref{eq:A*}, our model predicts the difference of $s_T$ due to different turbulence length scales.
Furthermore, the uncertainty range of model predictions in Fig.~\ref{fig:h2_phi} basically covers the deviation of $s_T$ in DNS results.
%
%

\begin{figure}
  \centering
  \includegraphics[width=80 mm]{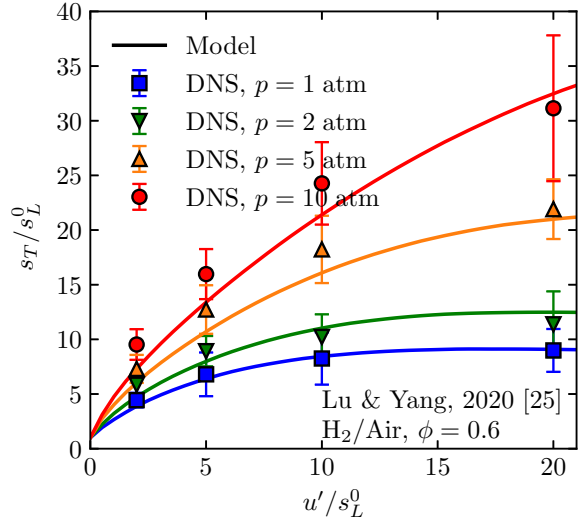}
  \caption{
    Comparisons of $s_T$ obtained from the DNS~\cite{Lu_2020} (symbols with error bars for one standard deviation) and the proposed model Eq.~\eqref{eq:st} (lines) for lean hydrogen flames at various pressures and $l_t/\delta_L^0=1$.
    }
  \label{fig:h2_pressure}
\end{figure}

Lu and Yang~\cite{Lu_2020} investigated the pressure effects on the model prediction $s_T$ for lean hydrogen premixed flames.
%
In the present work, the model of $s_T$ is further improved with the length scale effects in Eq.~\eqref{eq:A*} and a universal model of $\mathcal{C}_0$ for different fuels in Eq.~\eqref{eq:C0f}.
Figure~\ref{fig:h2_pressure} compares the model predictions (solid lines) and DNS results (symbols) of $s_T$ at four pressures from 1 to 10 atm.
%
The present model keeps the good performance~\cite{Lu_2020} on predicting $s_T$ in a broad range of pressure conditions.

\subsection{Methane}

Methane is the largest component of natural gas, which is one of the major energy sources.
The methane/air mixture has Le close to 1, resulting in $I_0\approx 1$. Hence, the turbulent burning velocity is mainly controlled by the flame area ratio in Eq.~\eqref{eq:st_IA}.

\begin{figure}
  \centering
  \includegraphics[width=80 mm]{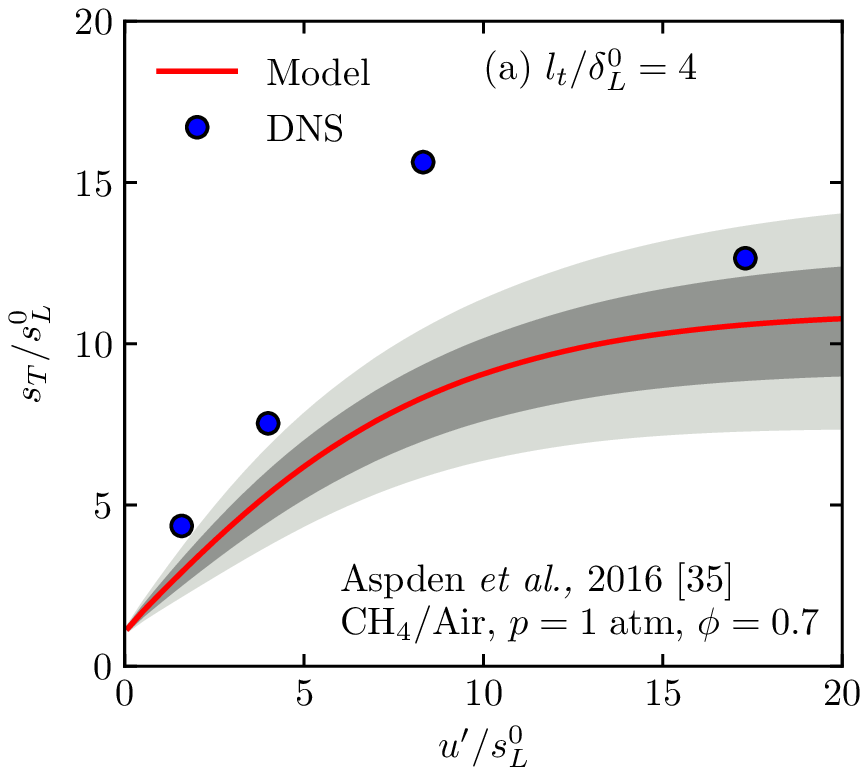}
  \includegraphics[width=80 mm]{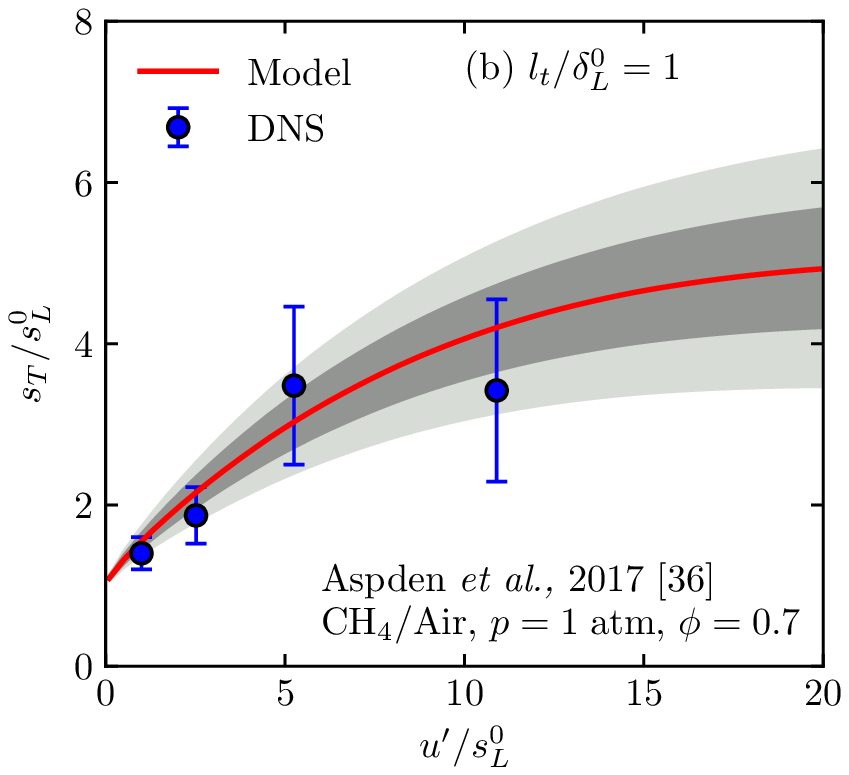}
  \caption{
    Comparisons of $s_T$ obtained from the DNS (symbols with error bars for one standard deviation) and the proposed model Eq.~\eqref{eq:st} (lines) for methane flames at $p=1$ atm and $\phi=0.7$. (a) Cases~\cite{Aspden_2016} with $l_t/\delta_L^0=4$; (b) cases~\cite{Aspden_2017} with $l_t/\delta_L^0=1$.
    Dark and light shades denote one and two standard deviations for the model uncertainty range, respectively.
    }
  \label{fig:ch4_dns}
\end{figure}

In Fig.~\ref{fig:ch4_dns}, the model predictions of $s_T$ (solid lines) are assessed by the DNS results (symbols) of lean methane premixed flames with $\phi=0.7$, $p=1$ atm, and different length ratios $l_t/\delta_L^0=4$ and $l_t/\delta_L^0=1$ from Aspden \textit{et al}.~\cite{Aspden_2016,Aspden_2017}.
%
%
The DNS results indicate that the turbulent burning velocity increases with $l_t/s_L^0$. In general, $s_T/s_L^0$ with $l_t/\delta_L^0=4$ is about two times of that with $l_t/\delta_L^0=1$, consistent with the scaling in Eq.~\eqref{eq:l_scaling}.
This difference is well captured by the model by considering the length scale effect in Eq.~\eqref{eq:A*}.
%
%

\begin{figure}
  \centering
  \includegraphics[width=80 mm]{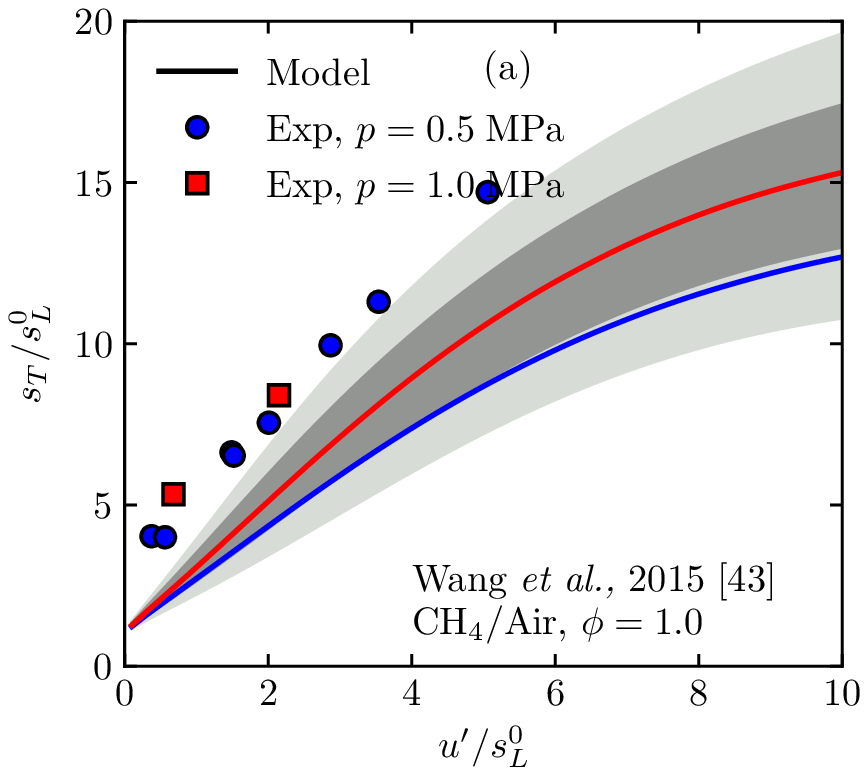}
  \includegraphics[width=80 mm]{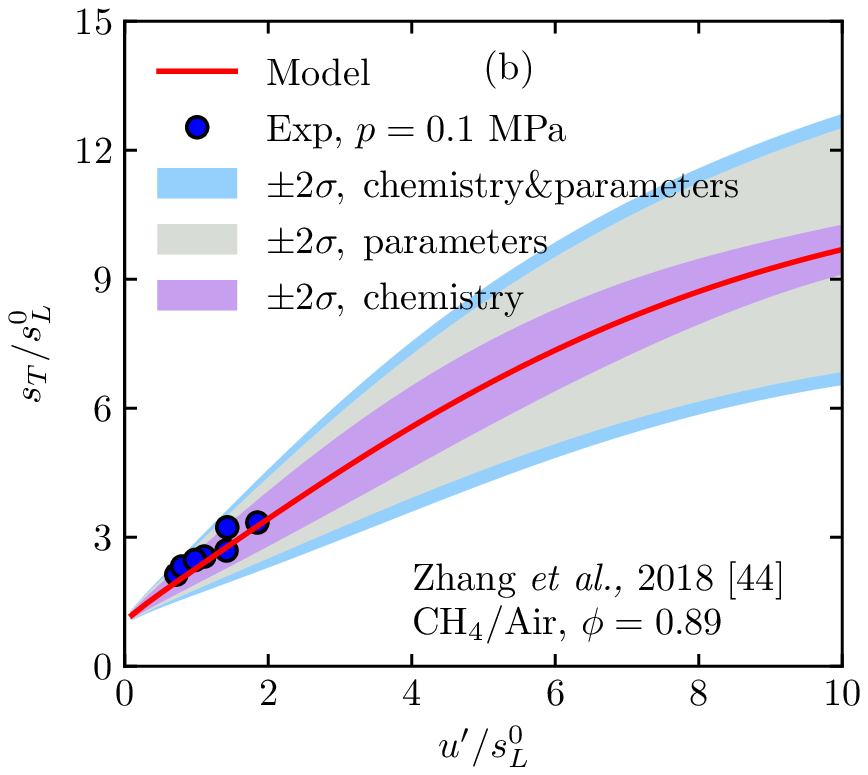}
  \caption{
    Comparisons of $s_T$ obtained from the experiment (symbols) and the proposed model Eq.~\eqref{eq:st} (lines) for methane flames at different pressures.
    (a) Cases~\cite{Wang_2015} with $\phi=1.0$, $p=0.5$, and 1.0 MPa.
    Dark and light shades denote one and two standard deviations for the model uncertainty range, respectively.
    (b) Cases~\cite{Zhang_2018} with $\phi=0.89$ and $p=0.1$ MPa.
    Purple, grey, and blue shades are the 95\% confidence intervals ($\pm 2\sigma$) considering the uncertainties of chemical kinetics, model parameters, and both ones, respectively.
    }
  \label{fig:ch4_p}
\end{figure}

In a series of experiments with a Bunsen burner, Wang and co-workers~\cite{Wang_2015,Zhang_2018,Zhang_2020} measured $s_T$ of turbulent premixed flames with various fuels at elevated pressures.
The present model predicts a linear growth of $s_T$ at small and moderate turbulence intensities in Fig.~\ref{fig:ch4_p}, generally agreeing with the experimental result. 
For the methane flames, the pressure only has a minor influence on $I_0$, so the model predictions of $s_T/s_L^0$ are similar at different pressures.
%
Additionally, Fig.~\ref{fig:ch4_p}b presents modeling uncertainties from chemical kinetics (purple shade), model parameters (grey shade), and both ones (light blue shade). For this case, the uncertainty from model parameters plays a dominant role.
Note that the chemistry uncertainty can be more significant for lean H$_2$/air mixtures and high pressures~\cite{Smith_FFCM}, as indicated in Fig.~\ref{fig:chemuq}.

\begin{figure}
  \centering
  \includegraphics[width=80 mm]{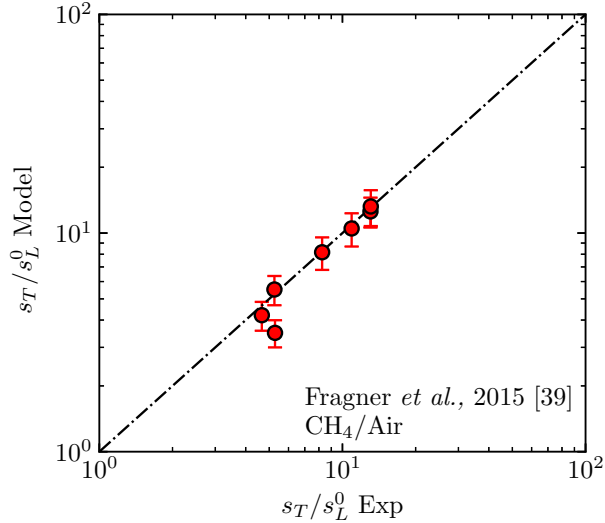}
  \caption{
    Comparisons of $s_T$ obtained from the experiment~\cite{Fragner_2015} and the proposed model Eq.~\eqref{eq:st} for methane flames at various conditions.
    Error bars denote one standard deviations for the model uncertainty range.
    }
  \label{fig:ch4_Fragner}
\end{figure}

Another dataset of $s_T$ for methane flames with pressure effects was obtained by Fragner \textit{et al}.~\cite{Fragner_2015} via a series of Bunsen flame experiments at different $p$ and $\phi$. %
As this dataset has roughly the same $u' \approx 0.44 \;\mathrm{m/s}$ and $l_t \approx 5.5 \;\mathrm{mm}$, Fig.~\ref{fig:ch4_Fragner} only plots model predictions versus experiment measurements instead of the plot $s_T/s_L^0$ versus $u'/s_L^0$.
As most points lie on the diagonal of the plot, the model predictions generally agree with the experimental results.
Similar to the previous validations, the model assessment for this dataset with $p = 0.1 \sim 0.4$ MPa and $\phi = 0.7 \sim 1.0$ confirms that the model in Eq.~\eqref{eq:st} is able to predict $s_T$ of methane flames at a wide range of conditions.

For most cases employed for the model validation, the model estimation of $s_T$ from Eq.~\eqref{eq:st} generally agrees with the DNS/experimental data.
Although the predictions are not perfectly accurate for some cases, they correctly capture the general variation trends of $s_T$.
Furthermore, the uncertainties in data statistics and model parameters are shown by error bars and shades in figures, respectively. We observe that the model prediction basically covers the data points with reasonable uncertainty ranges.

\subsection{Propane and mixed fuels}

Propane is a widely used alternative fuel for transportation.
The propane/air mixture can cover a range of Le by tuning the equivalence ratio, facilitating the investigation of thermal-diffusive effects.

\begin{figure}
  \centering
  \includegraphics[width=80 mm]{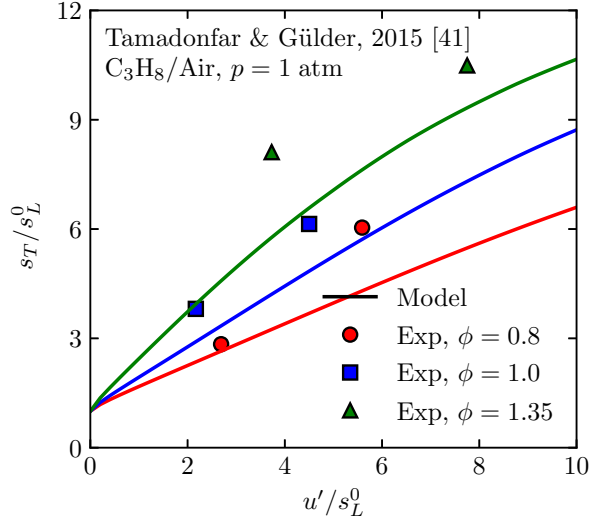}
  \caption{
    Comparisons of $s_T$ obtained from the experiment~\cite{Tamadonfar_2015} (symbols) and the proposed model Eq.~\eqref{eq:st} (lines) for propane flames  at $p=1$ atm and different equivalence ratios.
    }
  \label{fig:c3h8_Tamadonfar}
\end{figure}

\begin{figure}
  \centering
  \includegraphics[width=80 mm]{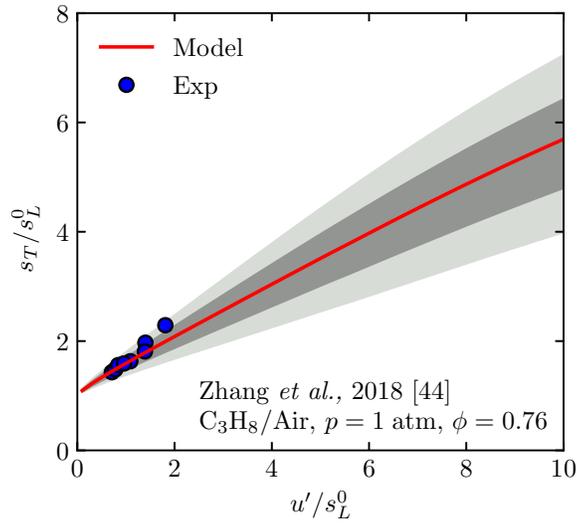}
  \caption{
    Comparisons of $s_T$ obtained from the  experiment~\cite{Zhang_2018} (symbols) and the proposed model Eq.~\eqref{eq:st} (lines)  for propane flames at $p=1$ atm.
    Dark and light shades denote one and two standard deviations for the model uncertainty range, respectively.
    }
  \label{fig:c3h8_Zhang}
\end{figure}

Figure~\ref{fig:c3h8_Tamadonfar} plots $s_T/s_L^0$ of propane/air Bunsen flames at three equivalence ratios in the experiment of Tamadonfar and G\"{u}lder~\cite{Tamadonfar_2015}. It shows that $s_T/s_L^0$ for the propane/air mixture increases with $\phi$, similar to the lean hydrogen flame in Fig.~\ref{fig:h2_phi}a. This effect of the equivalence ratio is captured by the proposed model.
For another experiment of propane/air Bunsen flames in Zhang \textit{et al}.~\cite{Zhang_2018}, Fig.~\ref{fig:c3h8_Zhang} compares the model prediction of $s_T$ against the experimental result.
The experiment data were obtained only at weak turbulence, and the present model predicts the linear growth of $s_T/s_L^0$ with $u'/s_L^0<2$.

%

%

\begin{figure}
  \centering
  \includegraphics[width=80 mm]{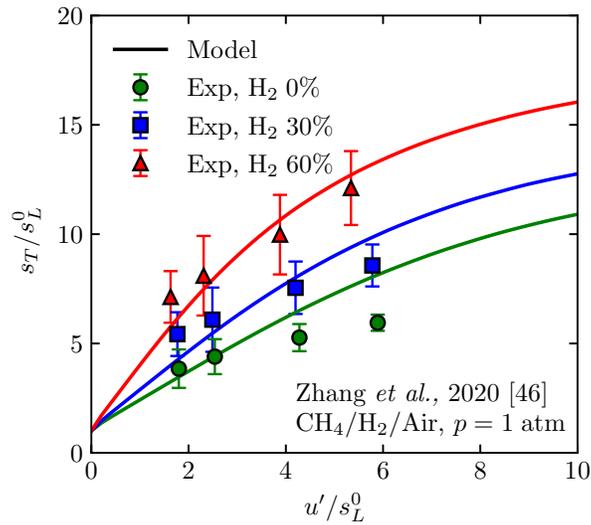}
  \caption{
    Comparisons of $s_T$ obtained from the  experiment~\cite{Zhang_2020} (symbols with error bars for one standard deviation) and the proposed model Eq.~\eqref{eq:st} (lines)  for methane/hydrogen flames with different hydrogen percentages in the mixed fuel.
    }
  \label{fig:ch4_h2}
\end{figure}

In practice, most fuels are mixtures of different components.
For specific applications, different fuels are mixed to adjust flame properties such as the laminar flame speed.
Among various mixed fuels, the hydrocarbon fuel blended with hydrogen has a wide range of Le, so it is particularly useful for the model validation with the effect of $\mathcal{C}_0$.
Zhang \textit{et al}.~\cite{Zhang_2020} studied the effect of differential diffusion on turbulent lean premixed flames of mixed fuels. In this experiment series, the extent of hydrogen enrichment for the CH$_4$/H$_2$/air mixture varies from 0\% to 60\%.
The laminar flame speed of each mixture is kept the same by adjusting $\phi$ in the experiment, whereas the measured turbulent burning velocities are different.
As shown in Fig.~\ref{fig:ch4_h2}, the mixture with more hydrogen extent has larger $s_T$ with stronger thermal-diffusive instability, and this trend is predicted by our model.

\subsection{Large hydrocarbon fuels}

Large hydrocarbon molecules can be found in many practical fuels such as gasoline, diesel, and jet fuels. They are also employed to construct the surrogate models for engineering applications.
For the large hydrocarbon fuels, the large Le reduces the stretched laminar burning velocity $s_L$, and the response of $s_L$ to flame stretch is generally not sensitive to pressure~\cite{Savard_2017}.

\begin{figure}
  \centering
  \includegraphics[width=80 mm]{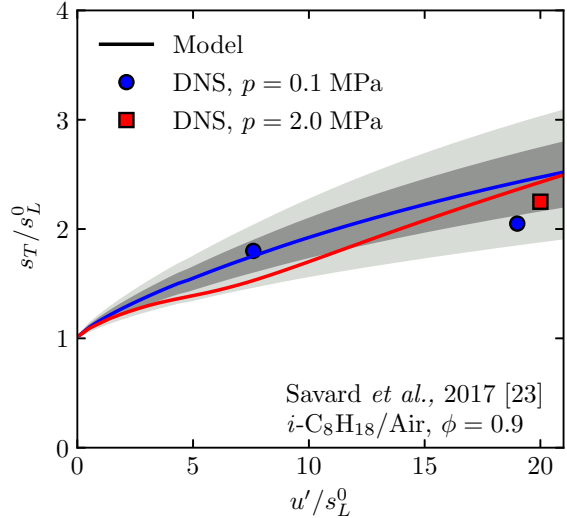}
  \caption{
    Comparisons of $s_T$ obtained from the DNS~\cite{Savard_2017} (symbols) and the proposed model Eq.~\eqref{eq:st} (lines) for \textit{iso}-octane flames with $\phi=0.9$, $p=0.1$ and 20 MPa.
    Dark and light shades denote one and two standard deviations for the model uncertainty range, respectively.
    }
  \label{fig:c8}
\end{figure}

\begin{figure}
  \centering
  \includegraphics[width=80 mm]{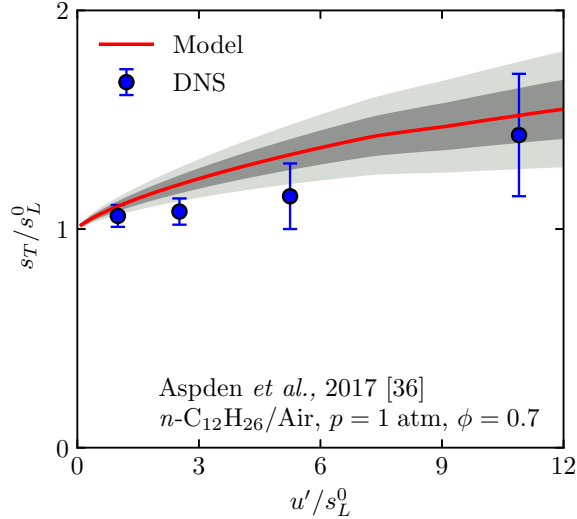}
  \caption{
    Comparisons of $s_T$ obtained from the  DNS~\cite{Aspden_2017} (symbols with error bars for one standard deviation) and the proposed model Eq.~\eqref{eq:st} (lines) for \textit{n}-dodecane flames  with $\phi=0.7$, $p=1$ atm.
    Dark and light shades denote one and two standard deviations for the model uncertainty range, respectively.
    }
  \label{fig:c12}
\end{figure}

Figure~\ref{fig:c8} compares model predictions (solid lines) and DNS results (symbols) of \textit{iso}-octane premixed flames~\cite{Savard_2017} at 1 and 20 atm.
The 143-species skeletal mechanism~\cite{Yoo_2013} was used in our calculation of $I_0$ for this case.
With the modeled $I_0$ pre-computed from laminar flames, the model correctly predicts that $s_T$ for heavy fuels is smaller than that for light fuels.
Furthermore, the model prediction is insensitive to pressure, consistent with the DNS observation.
Figure~\ref{fig:c12} validates the model by another DNS result for \textit{n}-dodecane flames~\cite{Aspden_2017}.
The small stretch factor for the large hydrocarbon fuel leads to a lower value of $s_T$, and the model prediction agrees well with the DNS result.

\subsection{Overall performance of model prediction}
\label{sec:performance}

In the present study, we assess the model Eq.~\eqref{eq:st} using 285 DNS and experimental results of $s_T$ over a wide range of fuels, equivalence ratios, pressures, turbulence intensities, and turbulence length scales.
The validations above have demonstrated that the proposed model gives overall good predictions of $s_T$.
%

\begin{figure}
  \centering
  \includegraphics[width=80 mm]{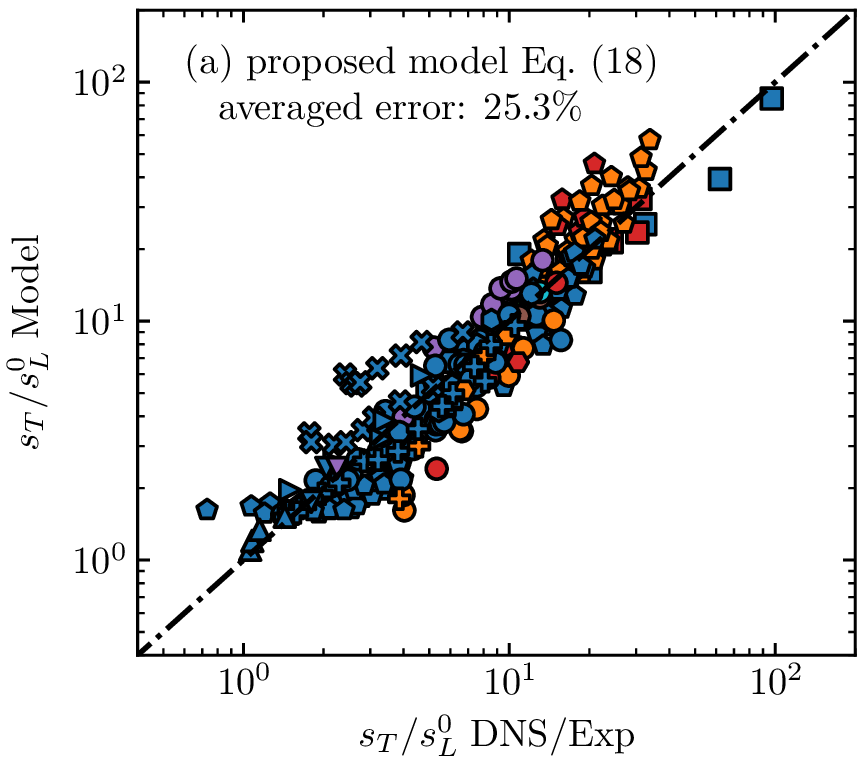}
  \includegraphics[width=80 mm]{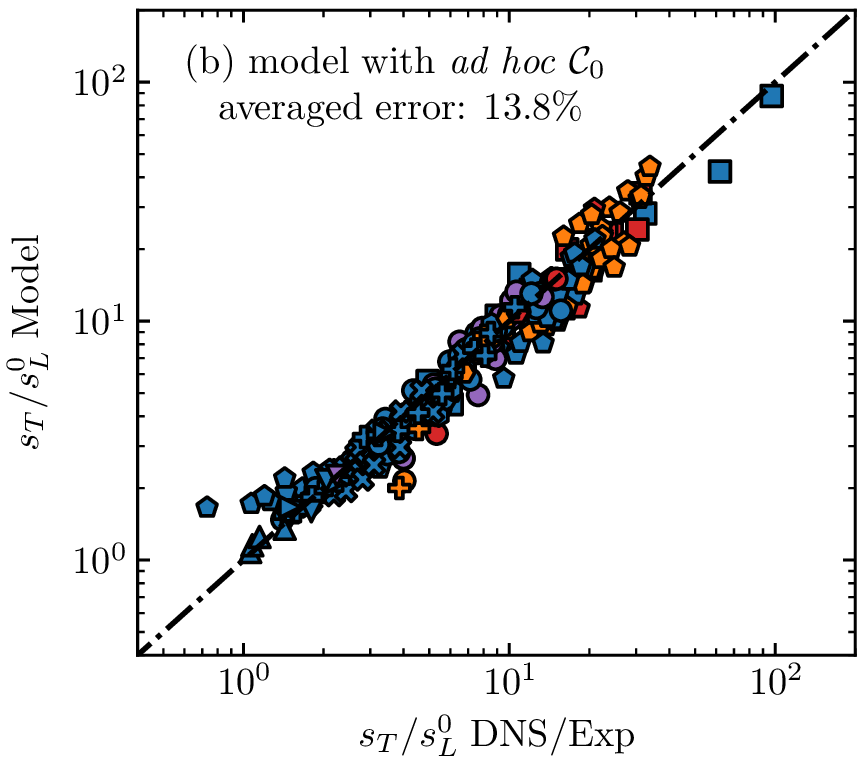}
  \caption{
    Comparison between DNS/experiment results and model predictions of $s_T/s_L^0$ for all the 285 data cases in Table~\ref{tab:cases}. (a) The proposed model Eq.~\eqref{eq:st}; (b) the model with \textit{ad hoc} $\mathcal{C}_0$.
    The symbol shape represents the fuel species, and the color denotes the pressure, which are the same with those in Fig.~\ref{fig:diagram}.
    }
  \label{fig:performance}
\end{figure}

Besides the model assessment through the representative cases for each type of fuels, Fig.~\ref{fig:performance} presents a comprehensive comparison between the model and DNS/experimental results for all the datasets listed in Table~\ref{tab:cases}.
Here, Fig.~\ref{fig:performance}a is obtained from the model in Eq.~\eqref{eq:st} without free parameters.
The corresponding averaged modeling error $\bar{\varepsilon}$ over all the cases is $25.3\%$. Here, the modeling error for each case is defined by
\begin{equation}
  \varepsilon = \dfrac{\lvert s_{T,\mathrm{model}}-s_{T,\mathrm{data}}\rvert}{s_{T,\mathrm{data}}}\times 100\%,
\end{equation}
where $s_{T,\mathrm{model}}$ is the $s_T$ predicted by the model in Eq.~\eqref{eq:st}, and $s_{T,\mathrm{data}}$ is the $s_T$ data obtained from DNS or experiment.

Considering the scattering of the DNS and experimental data across different groups and the intrinsic measurement and statistical uncertainties involved in the datasets, the averaged modeling error $25.3\%$ appears to be acceptable, and this performance is the best among existing $s_T$ models.
Here, the existing $s_T$ models are also assessed using the 285 cases in Table~\ref{tab:cases}. We find that the combination of sub models of the turbulence intensity, pressure, Lewis number, and fuel chemistry is important for the accurate prediction of $s_T$ over a wide range of conditions.
For example, the model~\cite{Muppala_2005} considering turbulence, pressure, and Lewis-number effects has relatively small $\bar{\varepsilon}=30\%$, and the models~\cite{Peters_2000,Yakhot_1988,Klimov_1983,Zimont_1979} only considering turbulence effects have $\bar{\varepsilon}$ larger than $50\%$.
%

On the other hand, the empirical model of the fuel-dependent coefficient in Eq.~\eqref{eq:C0f} can be improved, regarding the severe scatter of $\mathcal{C}_0$ in Fig.~\ref{fig:C0}.
If we apply $\mathcal{C}_0$ presented in Fig.~\ref{fig:C0} for the corresponding dataset, i.e., the model of Eq.~\eqref{eq:st} has one free parameter $\mathcal{C}_0$, the model prediction is significantly improved in Fig.~\ref{fig:performance}b, and $\bar{\varepsilon}$ is reduced from $25.3\%$ to $13.8\%$.
This large reduction demonstrates that the complexity of the fuel property and the hydrodynamic instability in weak turbulence is a major source of the uncertainties in the modeling of $s_T$.

Although the present model has been only validated for turbulent planar and Bunsen flames due to the requirement of the same consumption-based $s_T$ definition, it can be applied to flames with other geometries.
Our preliminary test shows that the present $s_T$ model prediction qualitatively agrees with the experimental data of V-flames of the methane/air mixture~\cite{Kheirkhah_2015,Smith_1979} and counterflow flames of the blended fuel mixture of CH$_4$/H$_2$ and C$_3$H$_8$/H$_2$~\cite{Abbasi-Atibeh_2019}, though different $s_T$ definitions were used in these flame data.
For spherical flames, most experiments observed the accelerating outwardly propagating flame front~\cite{Chaudhuri_2012,Shy_2015,Cai_2020} which has not reached a statistically stationary state, and $s_T$ was reported as an average over a range of flame radii~\cite{Wu_2015,Shy_2015,Cai_2020}.
Since the present model does not include this transient effect in turbulent flame development, it tends to overpredict $s_T$ for spherical flames.
Thus, the modeling of $s_T$ for the flames with complex geometries remains an open problem.


%

\section{Conclusion}
\label{sec:conclusion}

We propose a predictive model of the turbulent burning velocity for a wide range of conditions covering various fuels, equivalence ratios, pressures, and turbulence intensities and length scales.
Starting from the definition of the consumption speed, the model of $s_T$ involves the turbulence effects on the flame stretch and the flame area growth in Eq.~\eqref{eq:st_IA}.

The present model of $s_T$ in Eq.~\eqref{eq:st} has two major sub models.
First, the flame response under turbulence stretch is characterized by the stretch factor $I_0$ in Eq.~\eqref{eq:I0}, and $I_0$ is retrieved from a lookup table calculated from laminar counterflow flames in the implementation.  This model of $I_0$ incorporates the effects of the Lewis number and pressure for a variety of fuels.
Second, the flame area model~\cite{You_2020} based on Lagrangian statistics of propagating surfaces is improved to consider the effects of turbulence length scales and fuel characteristics. The scaling $(l_t/\delta_L^0)^{1/2}$ is incorporated to model the influence of turbulent diffusivity in Eq.~\eqref{eq:A*}.
An empirical model in Eq.~\eqref{eq:C0f} for the fuel-dependent coefficient $\mathcal{C}_0$ is proposed to quantify the effects of instabilities and fuel chemistry in weak turbulence.
%
In the implementation, $s_T$ is explicitly calculated from the algebraic model in Eq.~\eqref{eq:st} with several given reactant and flow parameters.
This model has no free parameter.


We perform a comprehensive validation for the $s_T$ model using 285 DNS/experimental cases reported from various research groups (see Table~\ref{tab:cases}).
The datasets for validation cover fuels from hydrogen to \textit{n}-dodecane, pressures from 1 to 20 atm, and lean and rich mixtures.
The model predictions and DNS/experimental results have an overall good agreement over the wide range of conditions, with the averaged modeling error 25.3\%.
Moreover, the model prediction involves the uncertainty quantification for empirical model parameters and chemical kinetic models.

The features of the present model are summarized as follows.
(1) The present model keeps the merit of previous ones~\cite{Lu_2020,You_2020} on predicting the bending phenomenon of $s_T$ via modeling competing mechanisms of growth and reduction of the turbulent flame area.
(2) The incorporation of the scaling for turbulence length scales extends the existing model to a wide range of turbulence parameters with $u'/s_L^0$ from 0.35 to 110 and $l_t/\delta_L^0$ from 0.5 to 80, so that the present model correctly predicts the growth of $s_T$ with $l_t/\delta_L^0$.
(3) 
Effects of detailed chemistry and transport are considered via the look-up table of $I_0$.
This sub model characterizes the thermal-diffusive effects of reactants on $s_T$, which is important to obtain correct bending curves of $s_T$ for thermal-diffusive unstable mixtures with Le $<1$.
(4) The fuel-dependent coefficient $\mathcal{C}_0$ describes the influence of instabilities in weak turbulence, which makes the present model applicable for various fuel mixtures.
%

We remark that the validation for the present $s_T$ model is restricted to planar and Bunsen flames. The model application for other flame geometries and different $s_T$ definitions requires further investigation.
Furthermore, this model still needs to be improved and extended for practical combustion problems such as the flame kernel development, swirling flame stabilization, and spray combustion.


\section*{Acknowledgement}

This work has been supported in part by the National Natural Science Foundation of China (Grant Nos.~91841302, 11925201, 11988102 and 91541204) and the Xplore Prize.



\bibliographystyle{elsarticle-num}
\bibliography{ST_Model}





\end{document}